\documentclass[a4paper]{article}
\usepackage{amsfonts}
\usepackage{epic}
\usepackage{epsfig}
\usepackage{subfigure}
\usepackage{amsmath}
\usepackage{amssymb}
\usepackage{rotating}
\DeclareMathOperator{\Ei}{Ei}
\begin{document}
\title{\bf Equilibrium configurations of tripolar charges}
\author{V. N. Yershov \\
{\small \it Mullard Space Science Laboratory} \\ 
{\small \it (University College London),} \\ 
{ \small \it  Holmbury St.Mary, Dorking RH5 6NT, UK} \\
{ \small vny@mssl.ucl.ac.uk}}
\date{}
\sloppy
\newcommand{\abs}[1]{\lvert#1\rvert}
\newcommand{\DoT}[1]{\begin{turn}{-180}\raisebox{-1.67ex}{#1}\end{turn}}
\newcommand{\up}{\upharpoonright}
\newcommand{\down}{\downharpoonright}
\newcommand{\NL}{\hspace{1.8ex}{\raisebox{1.5ex}
{\begin{rotate}{180}{$\nabla$}\end{rotate}}}}
\newcommand{\NR}{\hspace{1.4ex}\raisebox{-0.9ex}
{\begin{rotate}{63}$\nabla$\end{rotate}}}
\newcommand{\ND}{\hspace{-0.3ex}\raisebox{0.9ex}
{\begin{rotate}{-63}$\nabla$\end{rotate}}}
\newcommand{\NLP}{\hspace{1.8ex}\raisebox{1.5ex}
{\begin{rotate}{180}$\nabla$\end{rotate}}
\hspace{-2.35ex}\raisebox{0.25ex}{{\tiny +}}\hspace{0.5ex}}
\newcommand{\NRP}{\hspace{1.4ex}\raisebox{-0.9ex}
{\begin{rotate}{63}$\nabla$\end{rotate}}
\hspace{-1.8ex}\raisebox{0.25ex}{{\tiny +}}\hspace{0.6ex}}
\newcommand{\NDP}{\hspace{-0.3ex}\raisebox{0.9ex}
{\begin{rotate}{-63}$\nabla$\end{rotate}}
\hspace{0.1ex}\raisebox{0.25ex}{{\tiny +}}\hspace{0.5ex}}
\newcommand{\NLM}{\hspace{1.8ex}\raisebox{1.5ex}
{\begin{rotate}{180}$\nabla$\end{rotate}}
\hspace{-2.35ex}\raisebox{0.25ex}{\tiny {$-$}}\hspace{0.5ex}}
\newcommand{\NRM}{\hspace{1.4ex}\raisebox{-0.9ex}
{\begin{rotate}{63}$\nabla$\end{rotate}}
\hspace{-1.8ex}\raisebox{0.25ex}{\tiny {$-$}}\hspace{0.6ex}}
\newcommand{\NDM}{\hspace{-0.3ex}\raisebox{0.9ex}
{\begin{rotate}{-63}$\nabla$\end{rotate}}
\hspace{0.1ex}\raisebox{0.4ex}{\tiny {$-$}}\hspace{0.5ex}}
\newcommand{\NP}{\DoT{\DoT{$\nabla%
\hspace{-1.4ex}\raisebox{0.65ex}%
{\tiny{{$+$}}}$}}\hspace{0.4ex}}
\newcommand{\NN}{\DoT{\DoT{$\nabla%
\hspace{-1.4ex}\raisebox{0.7ex}%
{\tiny{{$-$}}}$}}\hspace{0.4ex}}
\newcommand{\DLP}{\hspace{-0.8ex}
{\DoT{\DoT{$\triangle$}}}\hspace{-1.5ex}\raisebox{0.3ex}
{\tiny \DoT{\DoT{$+$}}}\hspace{-1.75ex}\raisebox{0.1ex}
{\tiny \DoT{\DoT{$\blacktriangle$}}}
\hspace{0.2ex}}
\newcommand{\DRP}{\hspace{-0.8ex}
{\DoT{\DoT{$\triangle$}}}\hspace{-1.6ex}\raisebox{0.25ex}
{\tiny \DoT{\DoT{$+$}}}\hspace{-0.65ex}\raisebox{0.1ex}
{\tiny \DoT{\DoT{$\blacktriangle$}}}\hspace{0.1ex}}
\newcommand{\DUP}{\hspace{-0.8ex}
{\DoT{\DoT{$\triangle$}}}\hspace{-1.6ex}\raisebox{0.2ex}
{\tiny \DoT{\DoT{$+$}}}\hspace{-1.1ex}\raisebox{1.0ex}
{\tiny \DoT{\DoT{$\blacktriangle$}}}\hspace{0.5ex}}
\newcommand{\DLN}{\hspace{-0.8ex}
{\DoT{\DoT{$\triangle$}}}\hspace{-1.5ex}\raisebox{0.3ex}
{\tiny \DoT{\DoT{$-$}}}\hspace{-1.75ex}\raisebox{0.1ex}
{\tiny \DoT{\DoT{$\blacktriangle$}}}
\hspace{0.2ex}}
\newcommand{\DRN}{\hspace{-0.8ex}
{\DoT{\DoT{$\triangle$}}}\hspace{-1.6ex}\raisebox{0.25ex}
{\tiny \DoT{\DoT{$-$}}}\hspace{-0.65ex}\raisebox{0.1ex}
{\tiny \DoT{\DoT{$\blacktriangle$}}}\hspace{0.1ex}}
\newcommand{\DUN}{\hspace{-0.8ex}
{\DoT{\DoT{$\triangle$}}}\hspace{-1.6ex}\raisebox{0.2ex}
{\tiny \DoT{\DoT{$-$}}}\hspace{-1.1ex}\raisebox{1.0ex}
{\tiny \DoT{\DoT{$\blacktriangle$}}}\hspace{0.5ex}}
\newcommand{\DL}{\hspace{-0.8ex}
{\DoT{\DoT{$\triangle$}}}\hspace{-1.9ex}\raisebox{0.1ex}
{\tiny \DoT{\DoT{$\blacktriangle$}}}
\hspace{0.2ex}}
\newcommand{\DR}{\hspace{-0.8ex}
{\DoT{\DoT{$\triangle$}}}\hspace{-1.0ex}\raisebox{0.1ex}
{\tiny \DoT{\DoT{$\blacktriangle$}}}\hspace{0.1ex}}
\newcommand{\DU}{\hspace{-0.8ex}
{\DoT{\DoT{$\triangle$}}}\hspace{-1.5ex}\raisebox{1.0ex}
{\tiny \DoT{\DoT{$\blacktriangle$}}}\hspace{0.5ex}}
\newcommand{\DN}{\hspace{-0.8ex}
{\DoT{\DoT{$\triangle$}}}\hspace{-1.6ex}\raisebox{0.25ex}
{\tiny \DoT{\DoT{$-$}}}\hspace{0.1ex}}
\newcommand{\DP}{\hspace{-0.8ex}
{\DoT{\DoT{$\triangle$}}}\hspace{-1.7ex}\raisebox{0.3ex}
{\tiny \DoT{\DoT{$+$}}}
\hspace{0.2ex}}
\newcommand{\PiP}{\hspace{-0.1ex}
{\DoT{\DoT{$\Pi$}}}\hspace{-2.5ex}\raisebox{1.5ex}
{\tiny \DoT{\DoT{$+$}}}
\hspace{1.0ex}}
\newcommand{\PiN}{\hspace{-0.1ex}
{\DoT{\DoT{$\Pi$}}}\hspace{-2.5ex}\raisebox{1.5ex}
{\tiny \DoT{\DoT{$-$}}}
\hspace{1.0ex}}
\newcommand{\PiPN}{\hspace{-0.1ex}
{\DoT{\DoT{$\Pi$}}}\hspace{-2.5ex}\raisebox{1.5ex}
{\tiny \DoT{\DoT{$\pm$}}}
\hspace{1.0ex}}
\newcommand{\AP}{\hspace{-0.1ex}
{\DoT{\DoT{\sf P}}}\hspace{-2.3ex}\raisebox{1.5ex}
{\tiny \DoT{\DoT{$+$}}}
\hspace{1.0ex}}
\newcommand{\AN}{\hspace{-0.1ex}
{\DoT{\DoT{\sf P}}}\hspace{-2.3ex}\raisebox{1.5ex}
{\tiny \DoT{\DoT{$-$}}}
\hspace{1.0ex}}
\newcommand{\ANindex}[1]{\hspace{0.7ex}
{\scriptsize \DoT{\DoT{\sf P}}}\hspace{-1.6ex}\raisebox{0.7ex}
{\tiny \DoT{\DoT{$-$}}}
\hspace{-0.01ex}{\tiny \DoT{\DoT{$_#1$}}}\hspace{-0.4ex}}
\newcommand{\APN}{\hspace{-0.1ex}
{\DoT{\DoT{\sf P}}}\hspace{-2.3ex}\raisebox{1.5ex}
{\tiny \DoT{\DoT{$\pm$}}}
\hspace{1.0ex}}
\newcommand{\CP}[1]{\hspace{-0.01ex}
{\DoT{\DoT{\sf C}}}\hspace{-2.4ex}\raisebox{1.1ex}
{\tiny \DoT{\DoT{$+$}}
\hspace{1.0ex}{\scriptsize \DoT{\DoT{$#1$}}}}\hspace{-0.9ex}}
\newcommand{\CN}[1]{\hspace{-0.1ex}
{\DoT{\DoT{\sf C}}}\hspace{-2.3ex}\raisebox{1.1ex}
{\tiny \DoT{\DoT{$-$}}
\hspace{1.0ex}{\scriptsize \DoT{\DoT{$#1$}}}}\hspace{-0.9ex}}
\newcommand{\CZ}[1]{\hspace{-0.1ex}
{\DoT{\DoT{\sf C}}}\hspace{-2.2ex}\raisebox{1.1ex}
{\tiny \DoT{\DoT{$\circ$}}
\hspace{1.1ex}{\scriptsize \DoT{\DoT{$#1$}}}}\hspace{-0.9ex}}
\newcommand{\CPM}[1]{\hspace{-0.1ex}
{\DoT{\DoT{\sf C}}}\hspace{-2.5ex}\raisebox{1.1ex}
{\tiny \DoT{\DoT{$\pm$}}
\hspace{1.1ex}{\scriptsize \DoT{\DoT{$#1$}}}}\hspace{-0.9ex}}
\newcommand{\CZindex}[1]{\hspace{0.7ex}
{\scriptsize \DoT{\DoT{\sf C}}}\hspace{-1.6ex}\raisebox{0.7ex}
{\tiny \DoT{\DoT{$\circ$}}
\hspace{-0.01ex}{\tiny \DoT{\DoT{$#1$}}}}\hspace{-0.4ex}}
\newcommand{\CPindex}[1]{\hspace{0.7ex}
{\scriptsize \DoT{\DoT{\sf C}}}\hspace{-1.9ex}\raisebox{0.7ex}
{\tiny \DoT{\DoT{$+$}}
\hspace{-0.01ex}{\tiny \DoT{\DoT{$#1$}}}}\hspace{-0.4ex}}
\newcommand{\CPMindex}[1]{\hspace{0.7ex}
{\scriptsize \DoT{\DoT{\sf C}}}\hspace{-1.9ex}\raisebox{0.7ex}
{\tiny \DoT{\DoT{$\pm$}}
\hspace{0.2ex}{\tiny \DoT{\DoT{$#1$}}}}\hspace{-0.4ex}}
\newcommand{\YR}{{\sf Y}\hspace{-2.2mm}\raisebox{1.6ex}{%
\begin{turn}{143.0}{\sf{\small \bf l}}\end{turn}}\hspace{0.6mm}%
\raisebox{-1.7ex}{}}
\newcommand{\YL}{{\sf Y}\hspace{-3.3mm}\raisebox{1.6ex}{%
\begin{turn}{-143.0}{\sf{\small \bf l}}\end{turn}}\hspace{1.6mm}%
\raisebox{-1.6ex}{}}
\newcommand{\YD}{{\sf Y}\hspace{-0.15cm}{
\begin{turn}{-360.0}{\sf{\tiny \bf l}}\end{turn}}\hspace{0.06cm}}
\newcommand{\bigweirdy}{
{\sf Y}\hspace{-3.05mm}\raisebox{3.2ex}{\begin{rotate}{-140}
{$^\curlywedge$}\end{rotate}}%
\hspace{0.6mm}\raisebox{1.1ex}{\begin{rotate}{-40}
{$^\curlyvee$}\end{rotate}}%
\hspace{-0.1mm}\raisebox{-0.3ex}{$_\curlywedge$}
}
\newcommand{\tinyY}{{\tiny {\sf Y}}}
\newcommand{\turnedtinyY}{\begin{turn}{-180}
\hspace{-0.3mm}\raisebox{-0.85ex}{{\tiny {\sf Y}}}
\end{turn}~}
\newcommand{\weirdy}{{\tiny
\raisebox{0.4ex}{{\sf Y}}\hspace{-1.7mm}%
\raisebox{3.7ex}{\begin{rotate}{-135}
{$^\curlywedge$}\end{rotate}}%
\raisebox{1.4ex}{\begin{rotate}{-40}
{$^\curlyvee$}\end{rotate}}%
\hspace{0.05mm}\raisebox{0.1em}{$_\curlywedge$}
}
}
\newcommand{\Rc}{\sf \Breve{R}}
\newcommand{\Gc}{\sf \Breve{G}}
\newcommand{\Bc}{\sf \Breve{B}}
\newcommand{\Cc}{\sf \Breve{C}}
\newcommand{\Yc}{\sf \Breve{Y}}
\newcommand{\Mc}{\sf \Breve{M}}
\newcommand{\Pc}{\sf \Breve{P}}
\newcommand{\Nc}{\sf \Breve{N}}
\maketitle

\begin{abstract}
It is shown that an ensemble of particles with 
tripolar (colour) charges will necessarily cohere
in a hierarchy of structures, from simple clusters 
and strings to complex aggregates and cyclic 
molecule-like structures.
The basic combinatoric rule remains essentially 
the same on different levels of the hierarchy, 
thus leading to a pattern of resemblance between 
different levels. 
The number of primitive charges in each
structure is determined by the symmetry 
of the combined effective potential of this structure. 
The outlined scheme can serve as a framework for 
building a model of composite fundamental fermions. 
\end{abstract}
%
%
%
%
%
%
\section{Introduction}

It is known that the structures of important objects 
that physicists study, like stars, galaxies,
molecules, atoms, nucleons, and some particles, 
are equilibrium states between opposing 
forces of nature. Equilibrium
potentials are broadly used for modelling molecules 
\cite{lyubartsev00,burenin02}, vortices 
in superconductors \cite{sow98,wallraff03}, metal structures 
\cite{adams94}, and even granular materials \cite{blair03}.
Realistic interactions between molecules are known to have 
always attractive and repulsive components, due
to the fact that solids and liquids have the 
property of cohesion but, at the same time, do not
collapse to a point under the action 
of these forces. Such systems are modelled 
with potentials that comprise a repulsive inner 
and an attractive outer region (or vice versa).
A similar approach is often used in biochemistry \cite{leckband00}, 
colloid chemistry \cite{crocker96}, in material sciences
\cite{malescio02}, and many other branches of physics and 
chemistry.

In condensed matter, the interactions between neutral 
atoms are described by the equilibrium Lennard-Jones and Morse
potentials. The electron cloud of a neutral atom fluctuates about the
positively charged nucleus. The fluctuations in neighbouring 
atoms become correlated, inducing attractive dipole-dipole 
interactions. The equilibrium distance between two proximal 
atomic centres is determined by a trade-off between this 
attractive (van der Waals) dispersion force and a core-repulsion 
force that reflects electrostatic repulsion and the Pauli 
exclusion principle. 

For simplicity, the Lennard-Jones forces 
are typically modelled as effectively pair-wise additive,
and the velocities and positions of atoms are calculated by
numerical methods as a multi-body problem of mechanics.
The effective potential in these systems is  represented as 
a sum of one-body, two-body and three-body components.
The task can be simplified by coupling two-body and higher
multi-atom correlations in one model \cite{tersoff88}.
The central idea is that in real systems, the strength 
of each bond depends on the local environment, i.e. 
an atom with many neighbours forms weaker bonds than an 
atom with few neighbours. Then, one can use a pair potential, 
the strength of which depends on the environment (screened
potential in the Morse form). This is related to the exponential 
decay dependence of the electronic density and is 
usually written as: 

\begin{equation}
\begin{split}
& \sum_i E_i=\frac{1}{2}\sum_{i \neq j} V_{ij}, \\
& V_{ij}(r_{ij})=F_C(r_{ij})[F_\ominus(r_{ij})+b_{ij}F_\oplus(r_{ij})],
\end{split}
\end{equation}
where the potential energy is resolved into a site energy 
$E_i$ and a bonding energy $V_{ij}$ between the particles
$i$ and $j$; $r_{ij}$ is the distance between 
the particles (atoms); $F_{\ominus}$ and $F_{\oplus}$ are the
attractive and repulsive pair potentials:

\begin{equation}
\begin{split}
& F_{\ominus}(r)=~~a_\ominus\exp{(-\lambda_{\ominus}r)}, \\
& F_{\oplus}(r)=-a_\oplus\exp{(-\lambda_{\oplus}r)},
\end{split}
\end{equation}
and $F_C$ is a cut-off function. 
The strengths ($a_{\ominus}$ and $a_{\oplus}$) and the range of each bond 
depend on the local environment 
and are reduced when the number of neighbours is relatively high. 
This dependency is expressed by $b_{ij}$, which can enhance 
or diminish the repulsive force relative to the attractive 
force, according to the environment. 

In this paper we shall apply a similar approach to  the   
structures with tripolar charges,
taking into account the possibility of attractive and 
repulsive forces being different by their nature, rather than
both having an electrostatic origin. 
Indeed, in hadron and quark systems the attractive and
repulsive forces correspond to 
the strong (tripolar) interactions described by quantum 
chromodynamics \cite{suisso02}. 
Currently the attention of nuclear-physicists 
is focused primarily on the strong interactions in 
quark-gluon plasma and multi-quark systems. 
Some results of these studies, such as the estimation 
of the top-quark mass \cite{lep92} and prediction of 
pentaquarks \cite{diakonov97}, are confirmed 
by observations \cite{cdf95,do95,barmin03}, thus 
showing that the basic features of quantum chromodynamics 
are consistent with the subnucleonic reality. 

However, despite numerous 
publications on tripolar interactions, to date little attention
has been paid to the fact that the strong and electric charges  
can be modelled by two-component equilibrium fields, by analogy 
with molecular  and condensed matter physics. This approach can 
effectively yield new results. For example, the phenomenology of the hydrogen 
molecule (which is not yet well-un\-der\-stood)  has been recently
explained by introducing an attractive short-range Hulten (hadronic) 
potential between electrons, in addition to their conventional Coulomb 
repulsive potential \cite{santilli99}.  

Although the strong force {\it per se}  manifests both its 
attractive and repulsive nature \cite{eisberg85}, 
we shall include in our model both strong and electric field 
components. For the sake of simplicity 
we shall use identical particles, all having the same 
(unit) mass and charge. 

The tripolar fields are usually labelled with three primary
colours, which is also convenient for visualisation purposes
\cite{grutsch95}. 
For instance, a colour-neutral system (unaffected by any 
colour charge) can be represented (both mathematically and
graphically) as a superposition of 
three complementary colours in equal proportions 
(usually red, green and blue). This can be viewed as 
a ``white'' colour-charge (or ``black'', if the magnitudes 
of all three colours are mutually cancelled).

\section{Basic potential}

Let us consider a spherically symmetric equilibrium 
potential of a primitive particle {\sf P} with no properties, save 
its basic symmetry of SU(3)/U(1)-type. That is, this particle
has both electric and colour (unit) charges.
We shall regard a field $F(\rho)$ 
associated with such a particle as a superposition of two components, 
one attractive, $F_\ominus(\rho)$, and another repulsive, $F_\oplus(\rho)$,
satisfying the following conditions:
\begin{equation}
F(0)=F'(0)=0, 
\label{eq:pottozero}
\end{equation}
\begin{equation}
\exists \rho_\circ>0: \hspace{0.2cm}
F_\oplus(\rho_\circ) = -F_\ominus(\rho_\circ)
\label{eq:potreciprocal}
\end{equation}
(where $\rho$ is the radial coordinate).
We also assume the applicability of the 
least-action principle to the field $F$.
The condition (\ref{eq:potreciprocal}) implies that 
the components of the field cancel each other 
in the vicinity of some distance $\rho_\circ$, 
corresponding to equilibrium in a 
two-particle system. 

We suppose that both components of the field $F$ are  
closely related to each other (because they are underlied
by the same source -- the primitive particle {\sf P}).
This means that any local
changes in one component of the field are reflected in
the other, which would result in suppression of possible 
fluctuations in an equilibrium system composed of a few 
primitive particles.

In order to represent the colour-neutral systems we 
have to introduce a special notation for 
three colour polarities, complementary to each 
other. Let the vectors $\bf{r}$, $\bf{g}$, and $\bf{b}$
be the signatures of the three primary
colour charges (red, green and blue), such that
the ``white'' colour is 
\begin{equation}
\mathbf{w}=\mathbf{r}+\mathbf{g}+\mathbf{b},
\label{eq:primarycolours}
\end{equation}
where $\mathbf{w}$ is the diagonal of a unit matrix.
In order to satisfy (\ref{eq:primarycolours}), 
the $\bf{rgb}$-vectors could have the following 
components:
\begin{equation}
\begin{split}
\mathbf{r} & =(-1,+1,+1)^{\intercal} \\
\mathbf{g} & =(+1,-1,+1)^{\intercal} \\
\mathbf{b} & =(+1,+1,-1)^{\intercal}.
\end{split}
\label{eq:rgbcomponents}
\end{equation}
In the case of a system with mutually cancelled colour
charges we can write 
\begin{equation}
\mathbf{r}+\mathbf{g}+\mathbf{b}-\mathbf{w}=\mathbf{0},
\label{eq:rgbcancelled}
\end{equation}
which would correspond to a colour-neutral system with null electric charge.

With this notation, the field $F_i(\rho)$ of a particle ${\sf P}_i$ that has a colour
$\mathbf{c}_i\in\{\mathbf{r}, \mathbf{g}, \mathbf{b}\}$ can be written as
\begin{equation}
\mathbf{F}_i(\rho)=\mathbf{c}_iF_{\oplus}(\rho)+(\mathbf{c}_i-\mathbf{w}/3)F_{\ominus}(\rho).
\label{eq:csuperposition}
\end{equation}
In particular, in a system with $N=3$ complementary colour charges
(say, $\mathbf{c}_1=\mathbf{r}$,  $\mathbf{c}_2=\mathbf{g}$,
 $\mathbf{c}_3=\mathbf{b}$), the superposition of the fields $F_i$ will 
contain only the terms with $F_\oplus$:
\begin{equation*}
\sum_{i=1}^3 \mathbf{F}_i= \mathbf{w}F_\oplus,
\end{equation*}
because
\begin{equation*}
\sum_{i=1}^3(\mathbf{c}_i-\mathbf{w}/3)=\mathbf{0},
\end{equation*}
and the terms with $F_\ominus$ are mutually cancelled.

\begin{figure}[htb]
\centering
\begin{turn}{-90}
\epsfysize=2cm
\includegraphics[scale=0.7]{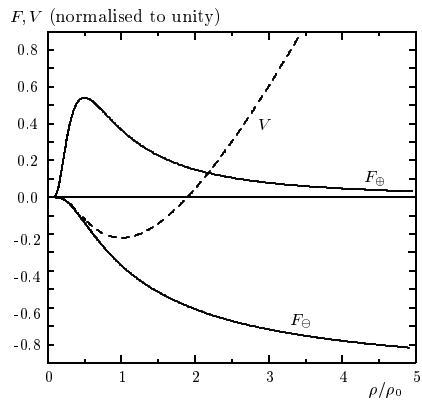}
\end{turn}
\caption{Components $F_{\ominus}$ and $F_\oplus$ of the equilibrium 
field $F$, and the corresponding potential, $V(\rho)$, for the signature 
$s_{ij}=-1$ in (\ref{eq:exp1}).} 
 \label{fig:twoexp}
\end{figure}

As a simple example of the split equilibrium field one can consider 
the field with the following components 
\begin{equation}
F_\ominus =F_0(\rho), \hspace{0.4cm} F_\oplus =F'_0(\rho),
\label{eq:sfield0}
\end{equation}
where
\begin{equation}
F_0(\rho)=s_{ij}a_0\exp(-\lambda_0\rho^{-1}).
\label{eq:exp1}
\end{equation}
The derivative in 
(\ref{eq:sfield0}) is
 taken with respect to the radial coordinate, $\rho$.
The coefficient $s_{ij}=\pm 1$ (signature) in 
(\ref{eq:exp1}) accounts for the sign 
of the interaction (repulsion or attraction) between two colour-charged 
particles, say,  ${\sf P}_i$ and ${\sf P}_j$. For the sake of simplicity
let the strength and range coefficients be normalised to unity 
($a_0=\lambda_0=1$). The functions $F_{\ominus}(\rho)$ and 
$F_{\oplus}(\rho)$,  
and the corresponding combined potential $V(\rho)$ are plotted 
in Fig.\ref{fig:twoexp}  (for $s_{ij}=-1$), where the 
unit distance, $\rho_\circ$, corresponds to (\ref{eq:potreciprocal}).

\section{Colour dipoles and tripoles}

Obviously, 
the simplest structures allowed by the tripolar 
field are the monopoles, dipoles and tripoles,
unlike the conventional bipolar (electric) field, which
allows only the monopoles and dipoles.
Here we shall consider the colour dipoles and tripoles.
The potentials shown in
Fig.\ref{fig:twoexp} correspond to a pair of like-charged 
($F_\oplus$ -- repulsive) primitive particles  
with unlike-colours ($F_\ominus$ -- 
attractive), which constitute a charged colour dipole
${\sf C}^2_{ij}={\sf P}_i+{\sf P}_j$. 
Here the indices $i$ and $j$ label the colour charges of the dipole's 
constituents: $i,j \in \{r, g, b\}$,  $i \neq j$; the upper
index ``2'' stands for the number of particles involved. 
As with any other dipole, the components of ${\sf C}^2_{ij}$ will 
oscillate near an equilibrium point at $\rho=\rho_\circ$, where
the potential $V(\rho)$ has a minimum.
The two components of $F$ are approximately antisymmetric in the vicinity of 
the origin, which would lead to suppression of these oscillations.
Then, the estimation of the ground-state 
energies (masses) of such a system will be simplified 
because one can neglect the oscillatory 
energy of ${\sf P}_i$ and ${\sf P}_j$ and, 
to a first-order approximation, 
compute the mass of the system as a sum of the masses of its constituents.

The existence of a second stationary
point in the potential -- at the origin -- means that the dipole's 
constituents, 
if confined within a very small volume,  
can be found in a spherically-symmetric superposition state at $\rho=0$. 
But this state is unstable and its spherical symmetry can be 
spontaneously broken, with $\rho\rightarrow \rho_\circ$,
resulting in the polarisation of the system.
This also breaks another fundamental symmetry -- that of 
scale invariance.  

Given the field $F(\rho)$ being split in two components, 
the rest energy of the particle {\sf P},
\begin{equation*}
- \int\limits_0^\infty F(\rho)d \rho,
\end{equation*}
can be resolved into two parts, containing
\begin{equation} 
 \int\limits_0^\infty F_{\ominus}(\rho)d \rho
\hspace{0.4cm} \text{and} \hspace{0.4cm}
\int\limits_0^\infty F_{\oplus}(\rho)d \rho,
\label{eq:energy1}
\end{equation}
which can be viewed as two mass terms,
$\tilde{m}_{\sf P}$ and $m_{\sf P}$, respectively.
With (\ref{eq:exp1}) normalised to unity, the second term, $m_{\sf P}$, 
is also a unity (let us denote this unit mass as $m_\circ$).   
But the first integral in (\ref{eq:energy1}) diverges 
($\tilde{m}_{\sf P}=\infty$), 
implying that within the chosen approach the primitive colour charges  
cannot exist in free states because of their infinite energies.
\begin{figure}[htb]
\centering 
\mbox{\hspace{-0.3cm}\subfigure[]{\begin{turn}{-90}\epsfig{figure=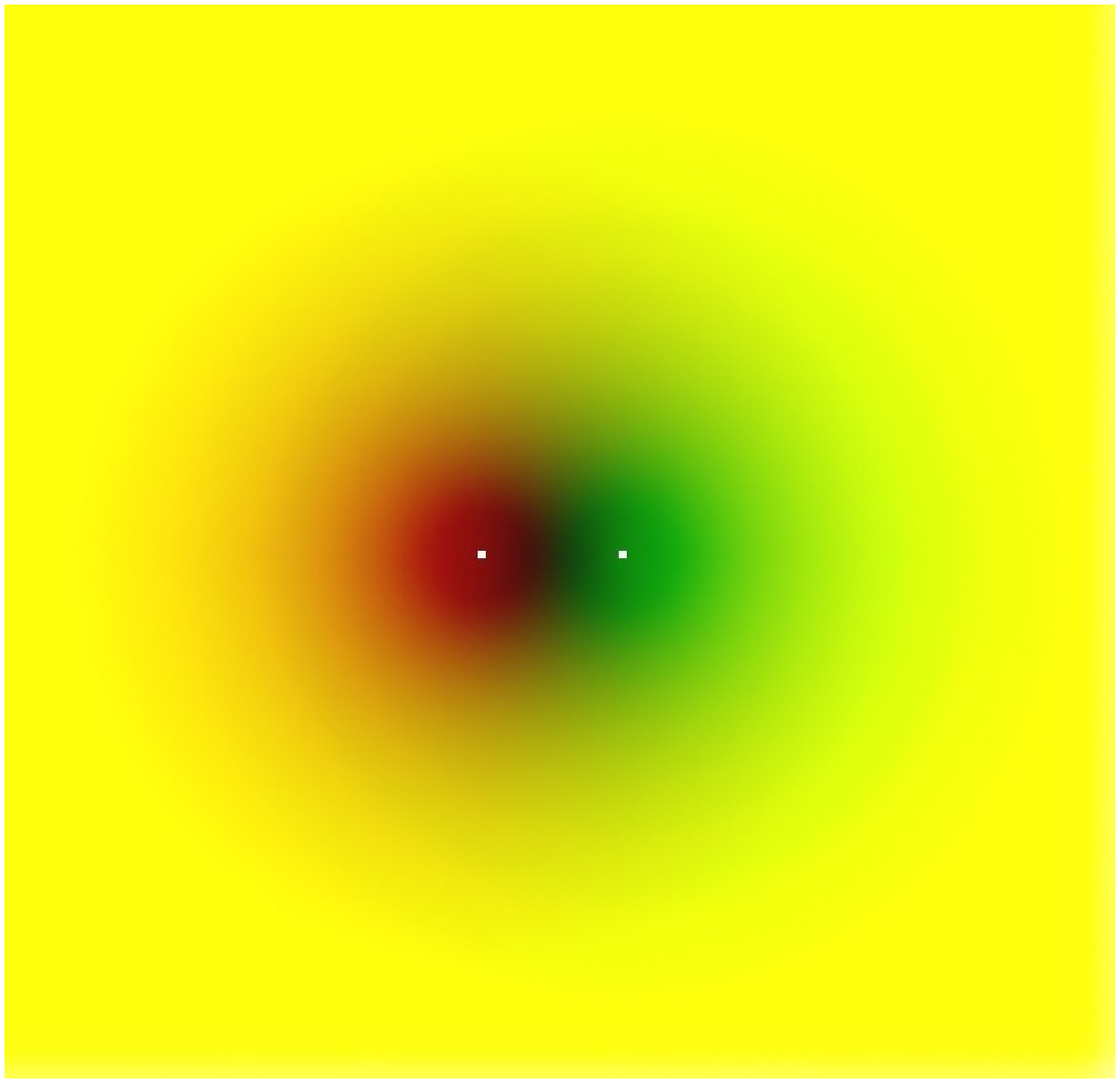, 
 width=4.0cm} \end{turn}}\quad
      \hspace{0.3cm}\subfigure[]{\begin{turn}{-90}\epsfig{figure=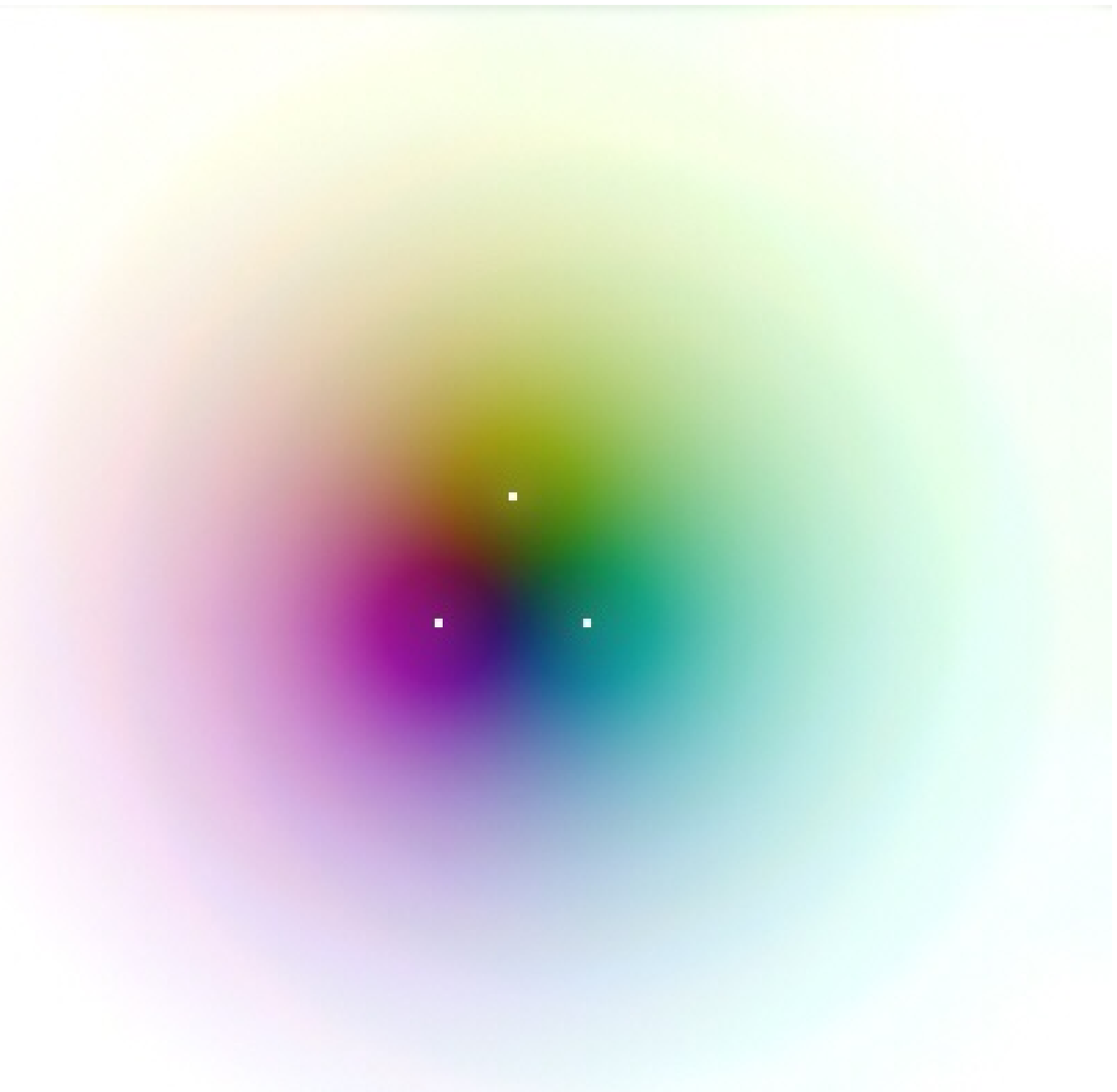,
 width=4.0cm}\end{turn}}}
\caption{(a): The field of the colour dipole, 
${\sf C}^2_{ij}$, is deficient in one colour, in this case blue, 
which is seen as an excess of the complementary colour (yellow) 
around the dipole. (b):
The field of the  $\triangle$-shaped (charged) tripole ${\sf C}^3$ in 
its equatorial plane. At some distance from the tripole its field 
is colourless (unaffected by any colour charge). 
The white dots mark the centres of the dipole and tripole constituents.
}
\label{fig:dipoletripolefield}
\end{figure}
The same is valid for the case of the colour-dipole, 
${\sf C}^2_{ij}$, which has only two of three possible 
colour fields $F_\ominus$ that cancel one another. This is illustrated in  
Fig.\ref{fig:dipoletripolefield}(a) where the dipole 
is shown for the colours $\mathbf{c}_i=\mathbf{r}$ and 
$\mathbf{c}_j=\mathbf{g}$. 
The deficient (diverging) colour, complementary to the other 
two, is blue, which is seen as an excess of yellow 
(white minus blue). 
Similar chromofields were discussed in \cite{martens03}, based 
on the Gaussian dielectric function and chiral 
chromodielectric model \cite{birse90, pirner92}, and also in
\cite{bowman04}.

Returning to the particle (inertial) masses, we must note that 
 in the current literature there is no agreement as to the origin 
of mass or inertia.
In the Standard Model of particle physics, the initially massless fundamental
particles acquire their masses through interactions with the Higgs
field. This is currently not yet supported by observations, and in this
paper we are free to adhere to a different view that mass is a purely 
electromagnetic phenomenon. In the simplified approach of this paper 
we shall not be considering any other forces rather than the electrostatic 
force caused by the equilibrium field $F(\rho)$. However, contrary to the 
conventional Coulomb gauge, we shall not 
regard the field $F(\rho)$ as acting instantaneously at a distance because 
this would be incompatible with the causality principle. It is more sensible 
to suggest that the field flow rate is not infinite. Then, there will be 
a time delay between the action on one part of a system
and the response from its another part. 
This can be viewed as inertia of the system, and the   
mass of such a system can be regarded as a measure of this delayed 
response to the external action. That is,
the more components of the system are to respond to this action
and the more mutually interacting components contribute to 
that response, the higher mass should be assigned to this system.    

To formalise the calculation of masses, we shall represent 
the discharge of the primitive colour particle with the use of 
auxiliary  $3\times 3$ singular matrices ~$\APN~_i$ containing 
the following elements:
\begin{equation}
^{\pm}p^i_{jk}=\pm\delta^i_j (-1)^{\delta^k_j},
\label{eq:pmatrix}
\end{equation}
where $\delta^i_j$ is the Kronecker delta-function;
the $\pm$-signs correspond to the sign of the charge; 
and the index $i$ stands for the colour ($i=1,2,3$ or red, green and blue).
The diverging components of the field can be represented by 
reciprocal elements:
\begin{equation*} 
\tilde{p}_{jk}=
p_{jk}^{-1}.
\end{equation*}
Then, we can define the charges and masses of the primitive 
particles by summation of these matrix elements:  
\begin{equation}
q_{\sf P}=\mathbf{w}^{\intercal}{\sf P}\mathbf{w} \hspace{0.1cm},
 \hspace{0.3cm} \tilde{q}_{\sf P}=\mathbf{w}^{\intercal} \tilde{\sf P}
\mathbf{w}
\label{eq:preoncharge}
\end{equation}
and
\begin{equation}
m_{\sf P}=\abs{\mathbf{w}^{\intercal}{\sf P}\mathbf{w}} \hspace{0.1cm},
\hspace{0.3cm} \tilde{m}_{\sf P}=\abs{\mathbf{w}^{\intercal} \tilde{\sf P}
\mathbf{w}}
\label{eq:mass2}
\end{equation}
($\tilde{q}_P$ and $\tilde{m}_P$ diverge). 
The same matrices {\sf P} can be used when calculating  
the signature $s_{ij}$ in (\ref{eq:exp1}) for the colours 
$i$ and $j$:
\begin{equation}
s_{ij}=-\mathbf{w}^\intercal{\sf P}_i {\sf P}_j\mathbf{w}.
\label{eq:seforce}
\end{equation}
In this notation the positively charged 
dipole ~${\CP{2}}$~ (Fig.\ref{fig:dipoletripolefield}a)
can be represented as a sum of two matrices, 
~$\AP_1$ and ~$\AP_2$:
\begin{equation}
{\CP{2}}=\AP_1+\AP_2=\begin{pmatrix} -1 & +1 & +1 \\ 
+1 & -1 & +1 \\ 0 & 0 & 0 
\end{pmatrix},
\label{eq:dipoleplus}
\end{equation}
with $q_{{\sf C}^2}=+2~[q_\circ]$.
If two components of the dipole are oppositely charged: 
\begin{equation}
{\CZ{2}}=\AP_1+\AN_2~
\label{eq:dipolezero}
\end{equation}
(of whatever colour combination), then
their electric fields cancel each other:
\begin{equation}
q_{{\CZindex{2}}} = 0
\label{eq:qcz}
\end{equation}
 implying also a negligibly small mass of this neutral dipole.
Of course,  the complete cancellation of the fields 
is possible only if the centres of both charges coincide; 
otherwise, the system is polarised  (as with any dipole). 
The degree of polarisation would depend on the distance 
between the components. Let us define the mass 
of a system containing, say, $N$ particles, as 
proportional to the number of these particles, 
wherever their field flow rates are not cancelled.
For this purpose, we shall consider (to a  first-order 
approximation) the total field
flow rate, $v_N$, of such a system as a 
superposition of the individual volume flow rates 
of its $N$ components. Then, the total mass can 
be calculated as the number
of particles, $N$, times the normalised to unity 
field flow rate $v_N$:
\begin{equation}
m_N=\abs{N v_N}.
\label{eq:mass}
\end{equation}
Here $v_N$ is computed recursively as a (Lorentz additive) 
superposition of the individual flow rates, $v_i$:
\begin{equation}
v_i=\frac{q_i+v_{i-1}}{1+\abs{q_i v_{i-1}}}, 
\label{eq:flowrate}
\end{equation}
where $i=2,\dots, N$; and $v_1=q_1$.
The normalisation condition (\ref{eq:flowrate}) 
expresses the common fact that the superposition
flow rate of, say, 
two antiparallel flows ($\uparrow \downarrow$) with equal
rate magnitudes 
$\abs{\mathbf{v}_\uparrow}=\abs{\mathbf{v}_\downarrow}=v$ vanishes 
($v_{\uparrow \downarrow}=0$), whereas, in the case of parallel
flows ($\uparrow \uparrow$) it cannot exceed the magnitudes of   
the individual flow rates ($v_{\uparrow \uparrow} \leq v$).
With this notation, the mass of, for instance,
the charged colour dipole will be: 
\begin{equation}
m_{\CPindex{2}} \approx 2, \hspace{0.2cm} \tilde{m}_{\CPindex{2}}=\infty.
 \label{eq:dubcharge}
\end{equation}
The neutral colour dipole will be massless: 
\begin{equation}
m_{{\CZindex{2}}} \approx 0
\label{eq:mcz}
\end{equation}
but still 
\begin{equation}
\tilde{m}_{{\CZindex{2}}}=\infty
\label{eq:mcz2}
\end{equation}
due to the null-elements in the matrix {\CZ{2}}~ (the dipole lacks, at
least, one colour charge to make it colour-neutral).
The infinities in (\ref{eq:dubcharge}) and (\ref{eq:mcz2}) imply that
neither {\CP{2}}~ nor {\CZ{2}}~ can exist in free states.
Of course, the flow rate of the electric field of the neutral dipole 
(and its corresponding mass) is cancelled only approximately 
(as with any dipole) because the centres of its constituents 
do not coincide. In an ensemble of a large number of 
neutral dipoles {\CZ{2}}~, not only electric but all 
the chromatic components of the field can be cancelled 
(statistically). 

Obviously, three complementary colour charges  
will tend to cohere and form a $\triangle$-shaped 
structure with the distance  $\rho_\circ$ (of equilibrium)  
between its components. Thus, 
by completing the set of colour-charges in the charged 
dipole (adding, for example, the blue-charged component 
to the system \CP{2}~ shown in Fig.\ref{fig:dipoletripolefield}a)
one would obtain a colour-neutral (but electrically 
charged) $\triangle$-shaped tripole: 
\begin{equation}
 {\sf C}^3:=[^{~g}_{r~b}]=\DL~~.
\label{eq:triangledef}
\end{equation}
Hereafter, we shall use the above triangular notation in the 
structural diagrams representing different tripole combinations
(one must not mistake these diagrams for algebraic equations).
The marked vertex of the triangle in (\ref{eq:triangledef})  
indicates one of the colour-charges, say, red, 
to visualise (in the structural diagrams) possible rotations of
tripoles with respect to each other. 
The positively charged $\triangle$-shaped tripole (\CP{3}~ or \DP),
which can be written in the matrix notation as 
\begin{equation*}
\CP{3}= \AP_1+\AP_2+\AP_3 = 
\begin{pmatrix} -1 & +1 & +1 \\ 
+1 & -1 & +1 \\ +1 & +1 & -1 
\end{pmatrix}~,
\end{equation*}
is colour-neutral at infinity but colour-polarised in the 
vicinity of its constituents (see Fig.\ref{fig:dipoletripolefield}b).
Both $m$ and $\Tilde{m}$ of ${\sf C}^3$ are finite: 
\begin{equation*}
m_\triangle=\tilde{m}_\triangle=3 ~[m_\circ],
\end{equation*}
since all of the diverging components in its combined chromofield 
are mutually cancelled (converted into the binding energy 
of the tripole).

\begin{figure*}[htb]
\centering 
\epsfig{figure=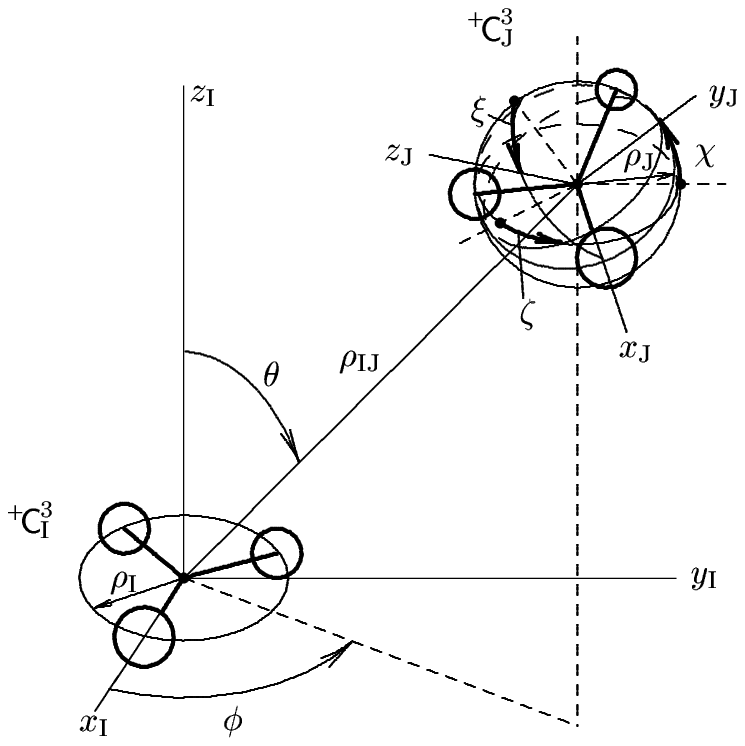,width=5.0cm}
\epsfig{figure=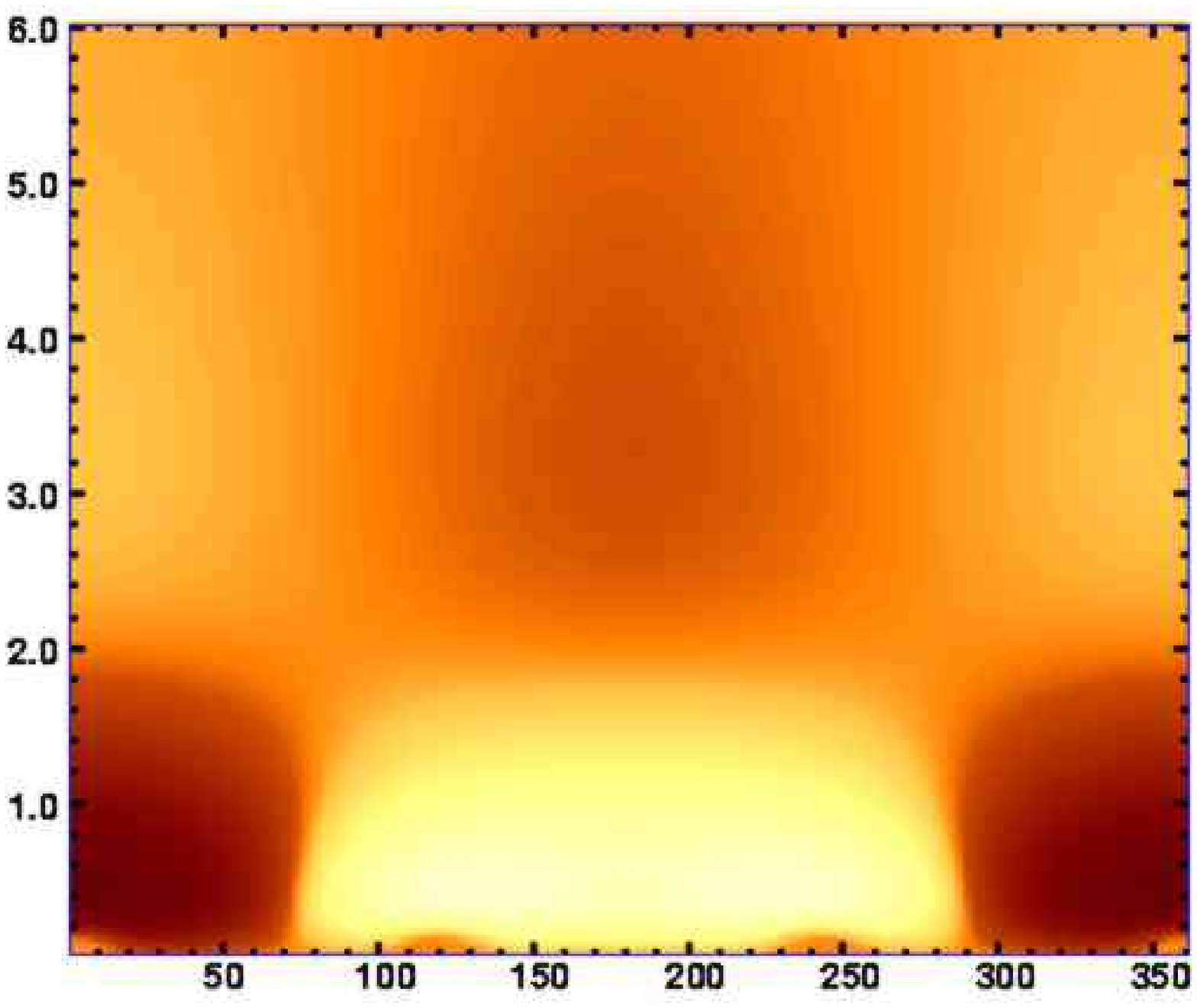,width=6.0cm}
 {\put(-170,155){
 \makebox(0,0)[t]{\footnotesize{$\frac{\rho_{\rm IJ}}{\rho_\circ}$}}}
 \put(-2,9){
 \makebox(0,0)[t]{\footnotesize{$^\circ~$}}}
 \put(-22,3){
 \makebox(0,0)[t]{\footnotesize{$\zeta~$}}}
 \put(-250,0){
 \makebox(0,0)[t]{\footnotesize{(a)}}}
 \put(-80,0){
 \makebox(0,0)[t]{\footnotesize{(b)}}}
 }
\caption{(a): Scheme for the computation of the potential energy
of two tripoles ${\sf C}^3_{\rm I}$ and 
${\sf C}^3_{\rm J}$ that 
form a charged doublet {\sf d}; and (b):
the potential energy of {\sf d} corresponding to  
$\phi,\theta, \chi, \xi=0$ and $\rho_{\rm I}=\rho_{\rm J}=
\rho_\circ/\sqrt{3}$. 
The position angle ($\zeta$) of 
${\sf C}^3_{\rm J}$ with respect to 
${\sf C}^3_{\rm I}$ 
is shown along the horizontal 
axis (in degrees); 
the vertical axis is the distance between two particles 
(in units of $\rho_\circ$). The darker regions correspond to 
lower energies.
}
\label{fig:yydoublet}
\end{figure*}

\section{Two-component systems of tripoles}

A part of the field of the tripole 
(in its equatorial plane)
is ring-closed \cite{kibler92}, 
whereas another part is extended (over the ring's poles). 
In its equatorial plane, the tripole possesses 
$2\pi \frac{m}{n}$-rotational 
symmetry ($m \leq n-1$, $n=2,3$) of  
the second- and third-order cyclic 
groups. With the dispersion of colour-charges, corresponding to 
this symmetry, different 
$\triangle$-tripoles can combine into chains.

A pair of tripoles would combine pole-to-pole with each
other forming a doublet ({\sf d}).
One can estimate the potential energy of this 
system by computing the pair-wise forces between its 
constituents. The scheme for this computation is shown 
in Fig.\ref{fig:yydoublet}(a).
The potential energy  of the doublet depends on the positions
of its components with respect to each other.
It can be computed as a superposition ($V_\Sigma$) of
two potentials, $V_{\ominus}=-\int{F_\ominus}$ and 
$V_{\oplus}=-\int{F_\oplus}$, based on 
the split field (\ref{eq:sfield0}). That is, 
\begin{equation}
\begin{split}
& V_{\ominus}(\rho_{ij})=s_{ij}\rho_{ij} \exp{(-\rho_{ij}^{-1})}+\Ei{(-\rho_{ij}^{-1})}, \\
& V_{\oplus}(\rho_{ij})=-s_{ij}\exp{(-\rho_{ij}^{-1})},
\end{split}
\label{eq:potential12}
\end{equation}
where $\rho_{ij}$ is the distance between the $i$-th charge
of the tripole~ $\CP{3}_{\rm I}$ and $j$-th charge of the 
tripole~ $\CP{3}_{\rm J}$; $i,j \in \{1,2,3\}$. 
For the sake of simplicity the arbitrary constants of integration 
in (\ref{eq:potential12}) are set
to zero. If we assume (also for simplicity) that the relative 
positions of the
primitive charges constituting the two tripoles of
the doublet are fixed within each of these tripoles,
then $V_\Sigma$ will have nine terms corresponding 
to the positions of the three charges of 
one tripole with respect to the three charges of the other:
\begin{equation}
V_\Sigma=-\mathbf{w}^\intercal V \mathbf{w}.
\label{eq:vsigma}
\end{equation}
Here the elements ${\rm v}_{ij}$ of the $3\times 3$ matrix $V$
are the following:
\begin{equation}
{\rm v}_{ij}=[V_{\ominus}(\rho_{ij})+V_{\oplus}(\rho_{ij})]
\hat{c}_{ik}\hat{c}^{kj},
\label{eq:velements}
\end{equation}
\begin{equation}
 \hat{c}_{ik}=c_{ik}+\delta_{ik}, 
\hspace{0.5cm} \hat{c}_{kj}=c_{kj}+\delta_{kj}
\label{eq:hatvelements}
\end{equation}
(repeated indices are summed over), where 
$c_{ik}$, $c_{kj}$ are the matrix elements of, respectively,~
 $\CP{3}_{\rm I}$ and~ $\CP{3}_{\rm J}$;
the distances $\rho_{ij}$ are those between the charges $i$ and 
$j$ belonging, respectively, to the tripoles~ 
$\CP{3}_{\rm I}$ and~ $\CP{3}_{\rm J}$; 
and the indices $i, j, k=1,2,3$ correspond to the three primary colours. 
The signature $s_{ij}$ in (\ref{eq:potential12}) is
computed with the use of (\ref{eq:seforce}). 
The Kronecker delta-symbols in (\ref{eq:hatvelements})
are added to the elements $c_{ik}$ and $c_{kj}$ to
satisfy the boundary condition of mutual cancellation (colourlessness)
of three complementary colour charges 
at infinity.
 
Besides its translational ($\mathbf{\rho}_{\rm IJ}$) and 
rotational ($\chi, \xi, \zeta$) degrees of freedom, 
the doublet {\sf d} has two degrees of freedom 
corresponding to the radial oscillations of its components
(radii $\rho_{\rm I}$ and $\rho_{\rm J}$). 
Making use of some obvious symmetries,  
we can reduce the dimensionality of the case (without loosing much 
information) by putting $\phi,\theta,\chi,\xi=0$ and
setting $\rho_{\rm I}$ and $\rho_{\rm J}$ to some fixed values, say, to 
the tripole's equilibrium radius ($\rho_\circ/\sqrt{3}$). 
The potential energy of this configuration is mapped 
in Fig.\ref{fig:yydoublet}(b) as a function 
of $\rho_{\rm IJ}$ and $\zeta$. 
It is seen that two like-charged tripoles can attract each other, if 
$\zeta\approx\pi$ and $\rho_{\rm IJ} > 2\rho_\circ$. 
The sign of the force between the tripoles depends 
on the position angle: for the angle  $\zeta=\pi/2$ 
the force is vanishing; it is attractive for 
$\pi/2 < \zeta \le 3\pi/2$:
\begin{equation}
\begin{split}
{\DUP} \rightarrow &  \leftarrow \DoT{\DUP} \\
{\DUP} \rightarrow &  \leftarrow~ {\DRP}  \\
{\DUP} \rightarrow & \leftarrow~  {\DLP}~
\end{split}
\label{eq:dipoleattractive}
\end{equation}
 and repulsive for $\abs{\zeta}<\pi/2$:
\begin{equation}
\begin{split}
\leftarrow~ {\DUP} \hspace{0.6cm} & \hspace{0.6cm} {\DUP} \rightarrow~ \\
\leftarrow~ {\DUP} \hspace{0.6cm} & \hspace{0.5cm}   \DoT{\DRP} \rightarrow \\
\leftarrow~  {\DUP} \hspace{0.6cm} & \hspace{0.5cm}  \DoT{\DLP}\rightarrow ~ .
\end{split}
\label{eq:dipolerepulsive}
\end{equation}
Thus, separated by distance  $\rho_{\rm IJ}>2\rho_\circ$ 
two tripoles can combine into the configuration  
\begin{equation}
{\sf d}^+={\DUP} \DoT{\DUP} \hspace{0.5cm} (\text{or}
\hspace{0.5cm} 
{\sf d}^- = \DUN \DoT{\DUN}~).
\label{eq:deltapm}  
\end{equation}
The existence of bifurcation points in the potential 
at $\rho_{\rm IJ}\approx 2\rho_\circ$
suggests the possibility of moving the system   
into a deeper potential well at $\rho \approx 0$ by squeezing 
it below  $\rho_{\rm IJ} = \rho_\circ$ (and keeping $\zeta=0$).

The width of the central potential well ($\zeta=\pi$, 
$\rho_{\rm IJ} > 2\rho_\circ$) 
 allows a certain degree of freedom for the 
constituents of {\sf d} to oscillate (rotating) 
within $\frac{2}{3}\pi < \zeta < \frac{4}{3}\pi$:
\begin{equation}
{\sf d}^+_\up={\DUP} \hspace{0.9ex} {\DRP}  \hspace{0.2cm} 
\rightleftarrows
\hspace{0.2cm} 
{\sf d}^+_\down = \DUP \hspace{0.9ex} {\DLP}~.
\label{eq:deltaprotation}  
\end{equation}
The strength and sign of the force between the components 
depends on $\zeta$. This implies the distance 
$\rho_{\rm IJ}$ being covariant with $\zeta$; that is, 
the translational and rotational oscillations of the doublet 
are synchronous. 
Note that in (\ref{eq:deltapm}) and (\ref{eq:deltaprotation}) we put 
the symbols~ \DP~  side-by-side, implying, however, 
that they rotate coaxially with respect to each other ($\theta, \xi=0$). 
The symbols $\up$ and $\down$ in (\ref{eq:deltaprotation}) 
denote the clockwise and anticlockwise rotations.
Over again, we would like to stress that 
the diagrams (\ref{eq:dipoleattractive})-(\ref{eq:deltaprotation})
express not otherwise than structural relationships
(like, for instance, the formulae in organic chemistry),
and by no means should they be mistaken for algebraic expressions. 
\begin{figure}[htb]
\centering
\epsfysize=2cm
\includegraphics[scale=0.32]{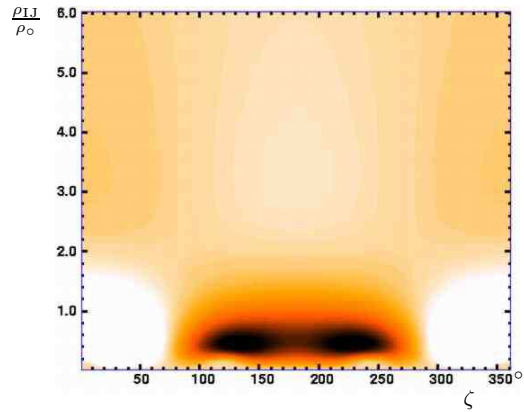}
 \put(-190,149){
 \makebox(0,0)[t]{\footnotesize{$\frac{\rho_{\rm IJ}}{\rho_\circ}$}}}
 \put(-2,11){
 \makebox(0,0)[t]{\footnotesize{$^\circ~$}}}
 \put(-21,4){
 \makebox(0,0)[t]{\footnotesize{$\zeta~$}}}
\caption{Potential of the tripole-antitripole system ${\sf d}^\circ$} 
 \label{fig:tantit}
\end{figure}

Similarly to the structure of the charged doublet, 
a pair of oppositely charged tripoles can form a neutral doublet
\begin{equation}
{\sf d}^\circ=\DUN \DoT{\DUP}~.
\label{eq:gamma}
\end{equation}
The corresponding potential 
is shown in Fig.{\ref{fig:tantit}. 
This system is massless, as well as colour-neutral 
(with $q_{{\sf d}^\circ}=0$, 
$m_{{\sf d}^\circ}=0$, and $\tilde{m}_{{\sf d}^\circ}=0$). 
Like ${\sf d}^+$ and ${\sf d}^-$, the neutral doublet
would oscillate, albeit with a smaller 
amplitude of its translational mode of oscillations
because its constituents are confined 
within a deep potential well at $\rho_{\rm IJ}\sim 0.5\rho_\circ$, 
as seen in Fig.{\ref{fig:tantit}. The rotational degree of freedom 
of the tripoles constituting the neutral doublet corresponds to
$\zeta=\pi \pm \pi/3$. 

The dynamics of the charged and neutral doublets deserves a more
detailed study but we shall address this problem elsewhere,
 since here we are interested mostly in reviewing the variety 
of possible particle configurations. 

\section{Three-component strings of tripoles}

A three-component string of $\triangle$-shaped tripoles 
[see Fig.\ref{fig:yyytriplet}(a)] have at least fifteen degrees 
of freedom (not taking into account possible flexions of the string). 
Making use of obvious symmetries we can reduce 
the number of parameters and analyse a basic 
case of an open string with three equidistant coaxial tripoles, 
$\triangle_{\rm I}$, $\triangle_{\rm J}$ and $\triangle_{\rm K}$. 
The potential of this system for the case 
$\rho_{\rm IJ}=\rho_{\rm JK}=\rho$ 
and $\zeta_{\rm K}=2\zeta_{\rm J}$ is shown in Fig.\ref{fig:yyytriplet}(b).
\begin{figure*}[htb]
\centering 
\epsfig{figure=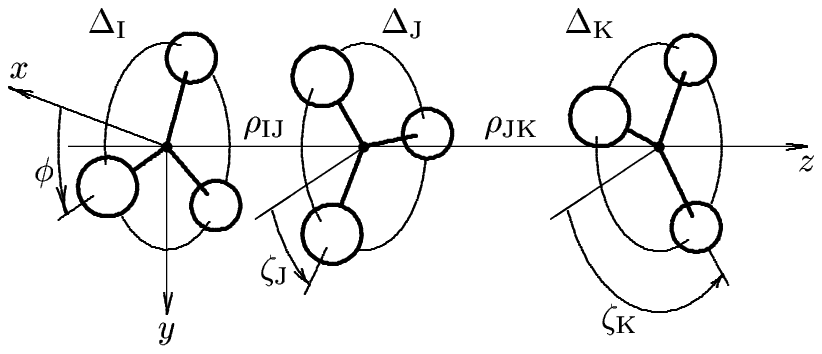,width=5cm}
  \hspace{0.5cm}
\epsfig{figure=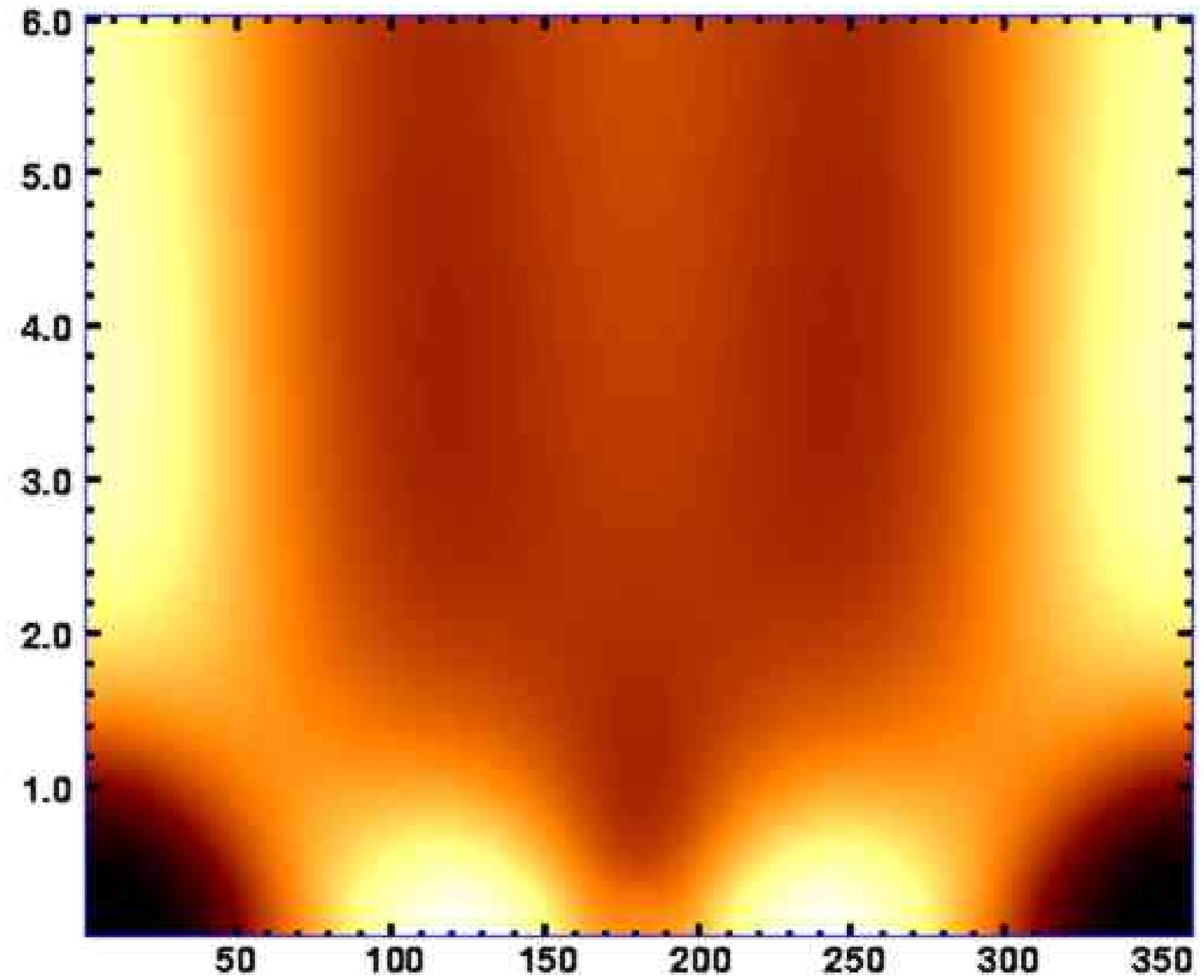,width=6cm}
 \put(-178,142){
 \makebox(0,0)[t]{\footnotesize{$\frac{\rho}{\rho_\circ}$}}}
 \put(-1,6.5){
 \makebox(0,0)[t]{\footnotesize{$^\circ~~\zeta_{\rm J}$}}}
 \put(-260,3){
 \makebox(0,0)[t]{\footnotesize{(a)}}}
 \put(-85,1){
 \makebox(0,0)[t]{\footnotesize{(b)}}}
\caption{(a): An open-string configuration of three coaxial 
tripoles $\triangle_{\rm I}$, $\triangle_{\rm J}$, $\triangle_{\rm K}$, 
and (b): its potential for the case  $\rho_{\rm IJ}=\rho_{\rm JK}=\rho$,
$\zeta_{\rm K}=2\zeta_{\rm J}$.}
\label{fig:yyytriplet}
\end{figure*}
\begin{figure*}[htb]
\centering 
\epsfig{figure=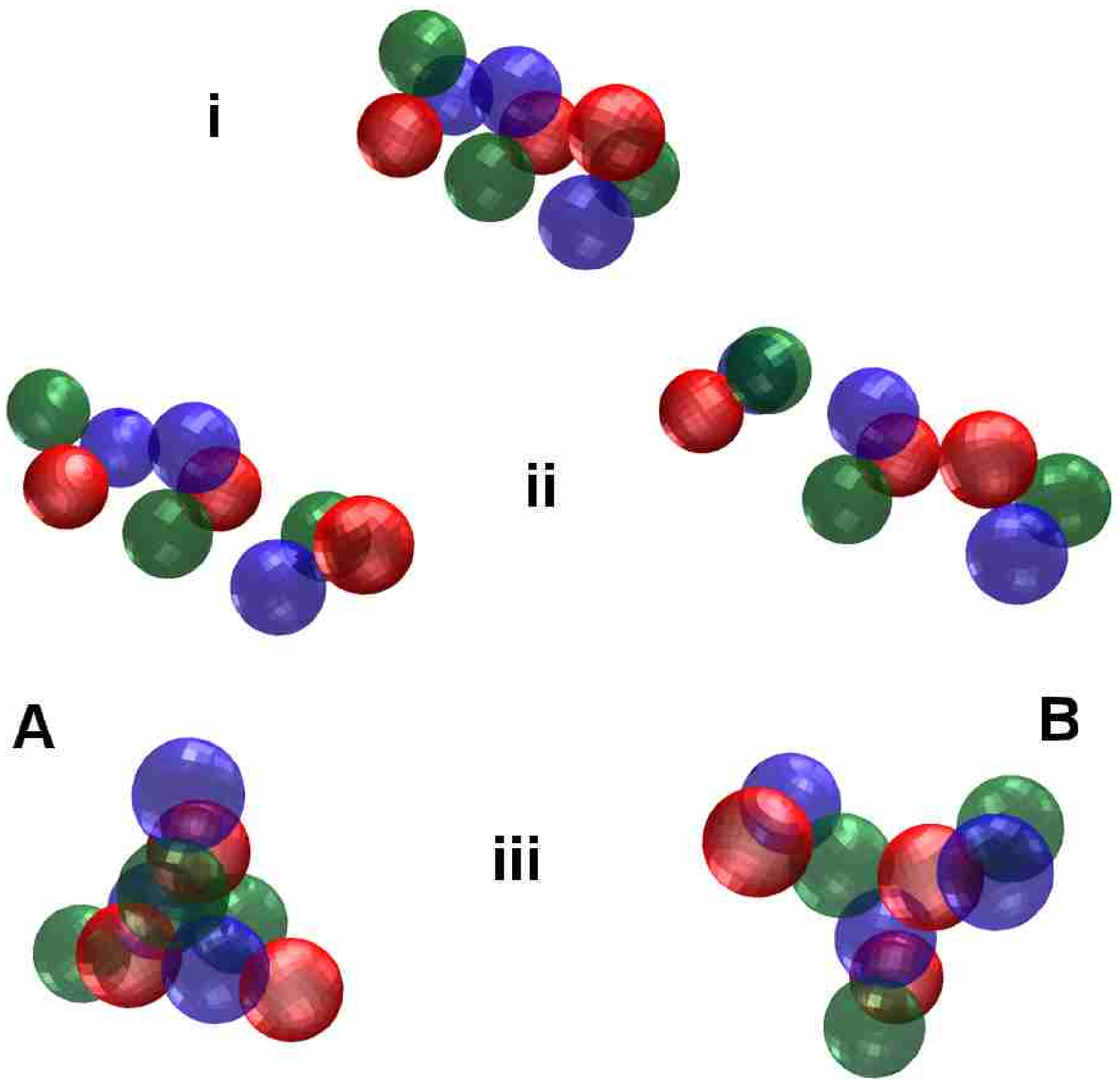,width=5.cm}
  \hspace{0.1cm}
\epsfig{figure=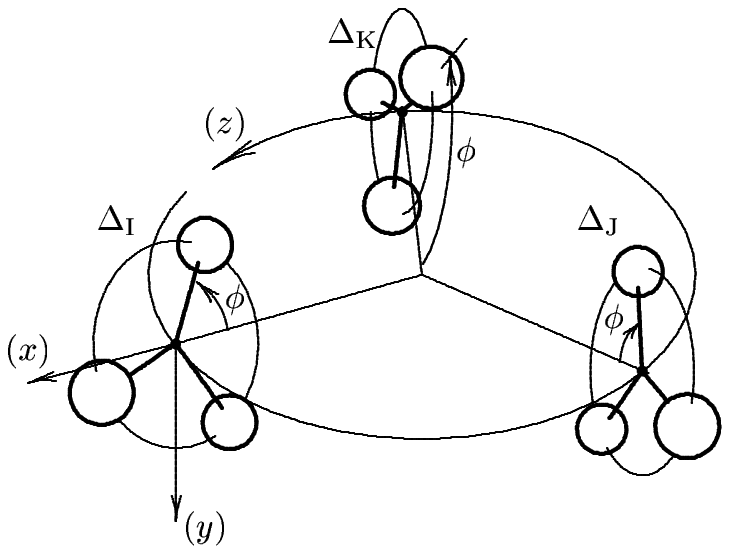,width=4.cm}
 \put(-200,2){
 \makebox(0,0)[t]{\footnotesize{(a)}}}
 \put(-40,2){
 \makebox(0,0)[t]{\footnotesize{(b)}}}
\caption{(a): Open string ({\rm i}) formed of three $\triangle$-tripoles is 
flexible: the loose ends of the string attract each other causing 
its closure in a loop. Two different bending
directions ({\rm ii}) correspond to two possible configurations 
of the loop ({\rm iii}), one with the vertices of its constituents
directed outwards from the loop (A) and another -- inwards (B).
(b): The scheme of the closed string. 
}
\label{fig:form3y}
\end{figure*}
\begin{figure*}[htb]
\centering 
\epsfig{figure=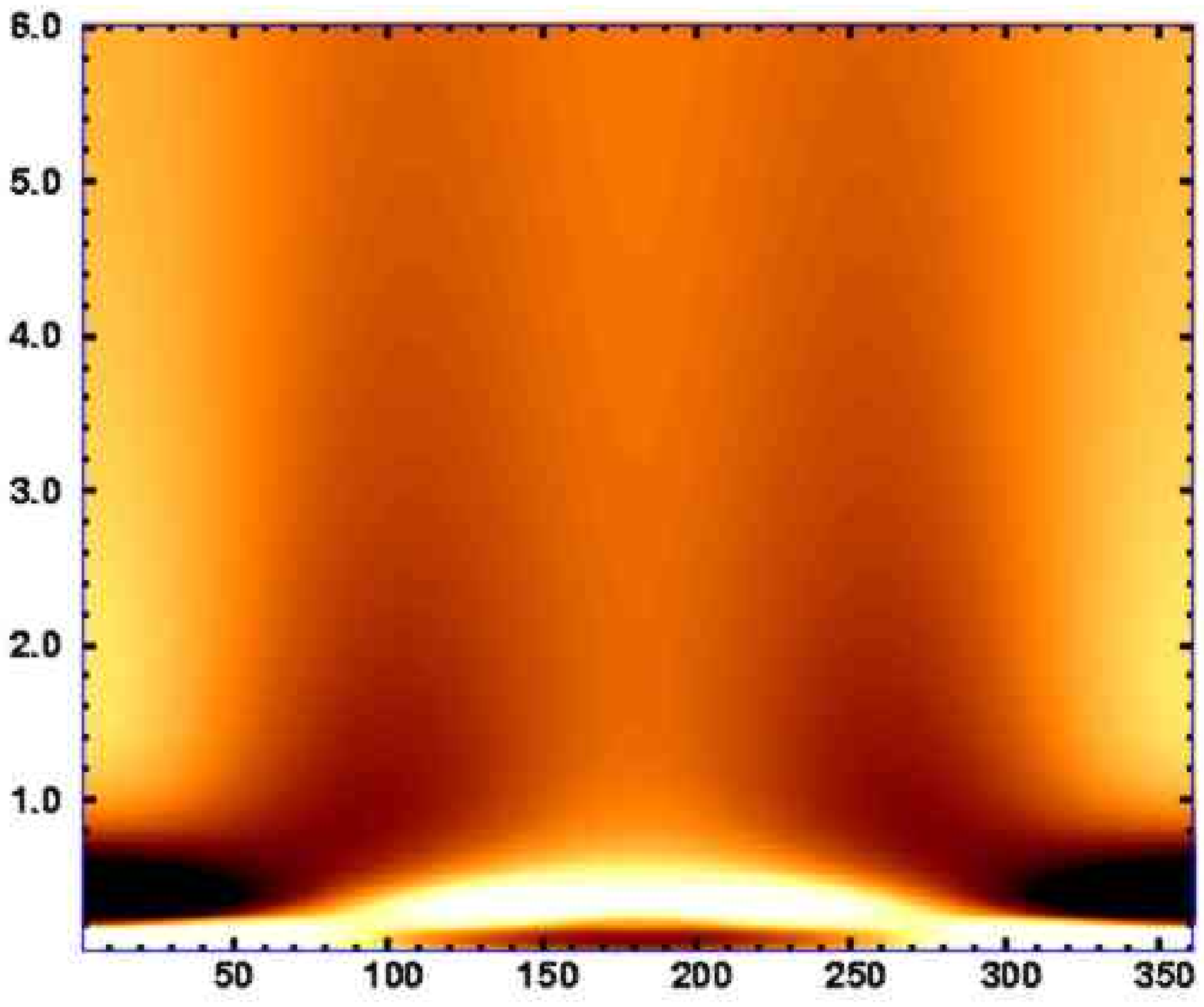,width=6.0cm}
 \put(-176,135){
 \makebox(0,0)[t]{
    \footnotesize{$\frac{\rho_{\rm Y}}{\rho_\circ}$}
                 }
               }
 \put(-7,8){
 \makebox(0,0)[t]{
   \footnotesize{$^\circ~~\zeta$}
                 }
           }
\quad
  \hspace{0.1cm}
  \epsfig{figure=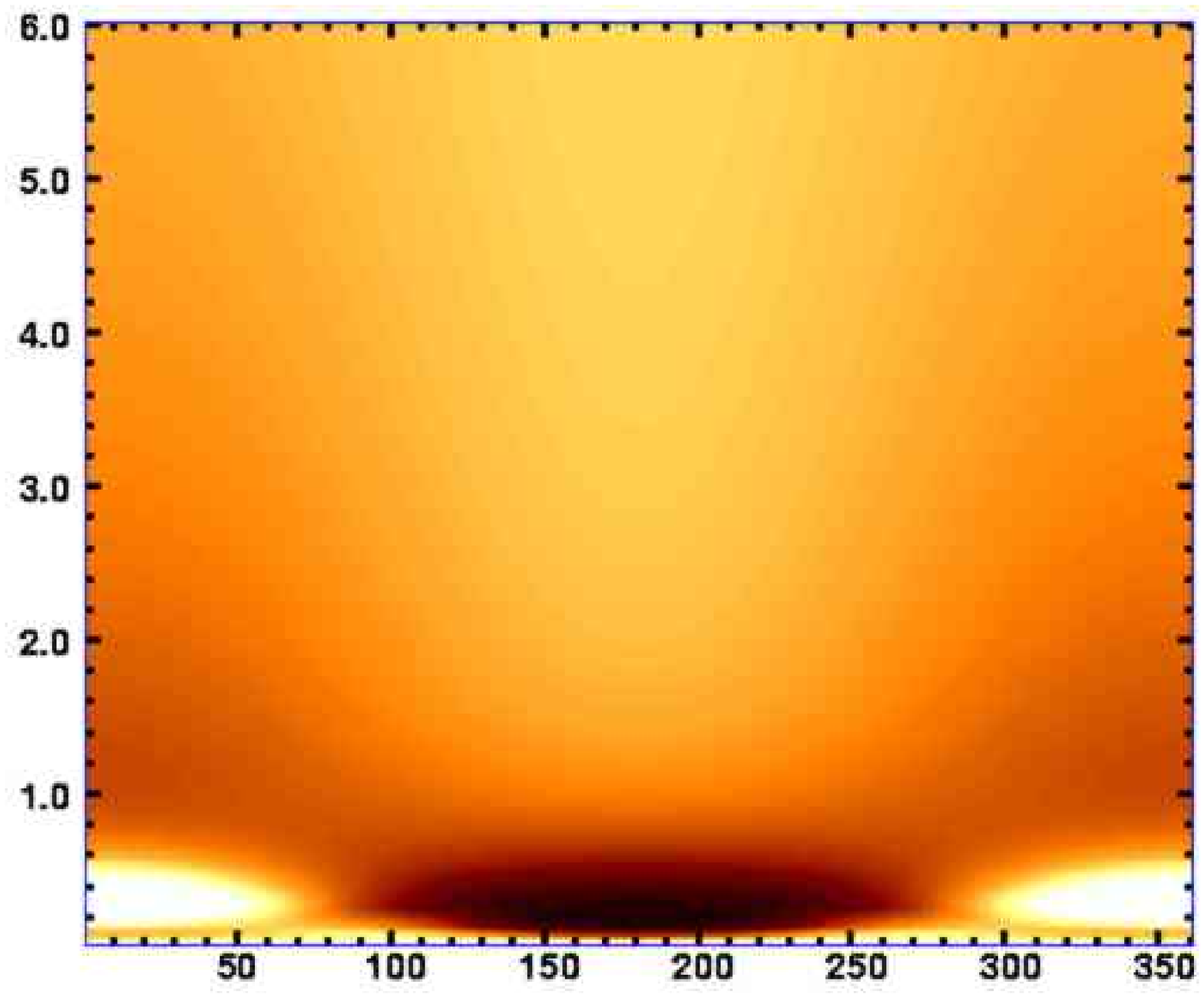,width=6.0cm}
 \put(-176,135){
 \makebox(0,0)[t]{
     \footnotesize{$\frac{\rho_{\rm Y}}{\rho_\circ}$}
                 }
               }
 \put(-6,9){
 \makebox(0,0)[t]{
       \footnotesize{$^\circ~~\phi$}
                 }
           }
 \put(-270,1){
 \makebox(0,0)[t]{\footnotesize{(a)}}}
 \put(-80,1){
 \makebox(0,0)[t]{\footnotesize{(b)}}}
\caption{
Potential of the three-component closed string ${\tt Y}$ 
 as a function: (a) of the position angle $\zeta$ between the 
$\triangle$-constituents of the string; and (b) of their phase 
angle $\phi$ for the fixed $\zeta=120^\circ$. 
The vertical axes in both plots correspond to radius 
$\rho_{\tt Y}$ (in units of $\rho_\circ$).
}
\label{fig:yyypotential}
\end{figure*}
 For $\rho > 2\rho_\circ$ there are two potential wells corresponding 
to  the position angle $\zeta=\pm \frac{2}{3}\pi$.  
Similarly to the doublet {\sf d}, the three-component 
string would perform translational and rotational oscillations 
which, however, cannot be stable because of various bifurcation 
points in its potential
(Fig.\ref{fig:yyytriplet}(b)). At the same time, one can see that the 
tripoles at the 
ends of the string attract each other, like those shown in 
the diagram (\ref{eq:dipoleattractive}),
and, due to the possible flexional 
deformations, the string {\DL}{\DU}{\DR} will necessarily close into 
a symmetric loop
\begin{equation}
{\tt Y}:=\raisebox{-1.4ex}{\DoT{\DU}}
\raisebox{1.5ex}{\DoT{\DL}}
\hspace{-0.5ex}
\raisebox{-1.5ex}{\DoT{\DR}} 
\label{eq:tty}
\end{equation}
as shown in Fig.\ref{fig:form3y}. This closure changes the 
properties of the system: in the case of the open string  
the variation of its phase angle $\phi$ [see  Fig.\ref{fig:yyytriplet}(a)] 
will not change the binding energy (relative distances) 
between its constituents, whereas the energy of the ring-closed
configuration, Fig.\ref{fig:form3y}(b), is phase-dependent.
Obviously, the energy states of the configurations {\sf A} 
($\phi=\pi$) and {\sf B} ($\phi=0$) in Fig.\ref{fig:form3y}(a) 
are distinct. In fact, these two configurations can be seen 
as $\pi$-phase-shifted states of the same structure, in which
its constituents, the $\triangle$-shaped tripoles, 
spin around its ring-axis. 
As in the case of the two-component system, the rotation
of tripoles around the ring-axis of ${\tt Y}$ implies
their circular translation along this axis, as well as   
the radial oscillations of ${\tt Y}$. 
This can be seen by analysing Fig.\ref{fig:yyypotential}, where the 
potential energy of ${\tt Y}$ is mapped as a function of 
the radius, $\rho_{\tt Y}$, position angle $\zeta$,  and phase 
angle $\phi$ between the components of the system. 

The potential wells in Fig.\ref{fig:yyypotential}(a) correspond
to the position angles $\zeta\approx \pm \frac{2}{3}\pi$
for a wide range of $\rho_{\tt Y}$.  
The charges spinning around the ring-closed axis 
of ${\tt Y}$ (clockwise or anticlockwise)  
will generate a toroidal (ring-closed) magnetic field which,
at the same time, will force these charges to move along the torus.
This  circular motion of charges will generate a secondary (poloidal)
magnetic field, contributing to the spin of these charges 
around the ring-axis, and so forth. The strength of the magnetic field
will be covariant with respect to $\rho_{\tt Y}$. The interplay of the 
varying toroidal and poloidal magnetic 
fields, oscillating $\rho_{\tt Y}$, and varying velocities of 
the rotating charges converts this system into a complicated 
harmonic oscillator with a series of eigenfrequences and oscillatory modes. 

The trajectories of charges (electric currents) in ${\tt Y}$, which are 
shown in Fig.\ref{fig:tripletcurrents},
\begin{figure}[htb]
\centering
\epsfysize=2cm
\includegraphics[scale=0.2]{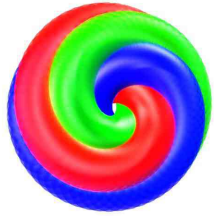}
\caption{Trajectories of the colour charges (electric currents)
in the structure ${\tt Y}$. The charges spin 
about the ring-axis of this structure and, at the same time,
synchroneously translate along this axis.
}
\label{fig:tripletcurrents}
\end{figure}
are helices with constant pitch and with two possible helical
signs: ${\tt Y}_\up$ (clockwise) or 
${\tt Y}_\down$ (anticlockwise), corresponding to 
two different signs of the internal angular momentum 
 of the constituents of this structure (around the 
ring-closed axis). Due to synchronisation of frequencies 
\cite{elnashar03}, by the closure of each $2\pi$-path along 
the ring-axis of ${\tt Y}$ the currents are additionally 
$\pi$-twisted about this axis. They need to travel twice 
along the ring to meet their initial phase condition 
({\sf A} or {\sf B}). The system looks much like a toroidal solenoid 
coil with winding number $w_{\tt Y}=3$.  

It is plain to see that if ${\tt Y}$ is positively charged, 
the vectors of its angular momentum and magnetic moment 
are always parallel (pointing in the same direction)
\begin{equation*}
\overset{\hspace{-2.0ex}\stackrel{{\tiny \ell}\rightarrow}
{\raisebox{0.1ex}{\tiny {\sf B}$\rightarrow$}}}{{\tt Y}^+}
\end{equation*}
whereas in the case of the negatively charged ${\tt Y}$ 
these vectors are antiparallel:     
\begin{equation*}
\overset{\stackrel{\hspace{-2.0ex}{\tiny \ell}\rightarrow}
{\raisebox{0.1ex}{\tiny $\leftarrow$ \hspace{-0.9ex}{\sf B}}}}{{\tt Y}^-}.
\end{equation*}
This property must be taken into account when considering interactions
between different {\tt Y}-particles.
The charge and mass of ${\tt Y}$ roughly correspond to the sum of the 
charges and masses of its nine constituents: $q_{\tt Y}=\pm 9$ 
(in units of $q_\circ$), $m_{\tt Y}=\tilde{m}_{\tt Y}=9$ (in units 
of $m_\circ$). At large distances from ${\tt Y}$ the combined 
chromofield of its constituents is colour-neutral and almost 
spherically symmetric. 
Thus, separated by large distances, 
${\tt Y}$-particles would behave as 
point-like colourless charges. At small distances from the 
particle one must take into account the chromatic polarisation 
of the field. 

\section{Chains of unlike-charged tripoles}

Hereafter -- in order to simplify our analysis -- we shall use 
the observation that $\triangle$-shaped tripoles, when clustered, 
form alike structures on different levels of complexity. For instance, 
a simple $\triangle$-tripole and the more complicated ring-closed 
triplet {\tt Y} (which consists of three $\triangle$-tripoles), 
both possess the same $\frac{2}{3}\pi$-rotational
symmetry. Then (up to a certain limit) one can use the same 
combinatoric rules when dealing with structures based on
{\tt Y}- and $\triangle$-particles. However, some new properties 
emerging on higher levels of complexity must be taken into account, 
such as, for example, the helicity property of {\tt Y}, which by no 
means can be found in $\triangle$. 
Based on the resemblance between different 
complexity levels, and using the pattern of attraction and 
repulsion between the tripoles with different positional angles 
[shown in the diagrams 
(\ref{eq:dipoleattractive}) and (\ref{eq:dipolerepulsive})] , 
let us explore the variety of possible structures based on 
the tripolar charges.
Of course, the detailed study of these structures 
and their properties need more rigorous numerical calculations 
and simulations, which we shall discuss elsewhere.

Based on the pattern (\ref{eq:dipoleattractive})
- (\ref{eq:dipolerepulsive}), one can find that
the unlike-charged doublets, 
${\sf d}^-$ and ${\sf d}^+$, can combine and form 
chains with two possible rotations of the components 
with respect to each other, clockwise:
\begin{equation}
{\sf d}^\circ_{2\up}= 
\DUN~ {\DRN}\hspace{-0.7ex}\DoT{\DRP} \DoT{\DUP}
\label{eq:dzeroup}
\end{equation}
or anticlockwise:
\begin{equation}
{\sf d}^\circ_{2\down}=
\DUN \hspace{0.7ex} \DLN \hspace{0.4ex} \DoT{\DLP} \hspace{0.2ex} \DoT{\DUP}~, 
\label{eq:dzerpdown}
\end{equation}
pole-to-pole to each other. The index ``2'' refers to the number 
of doublets involved (${\sf d}^-+{\sf d}^+ \rightarrow {\sf d}^\circ_2$).
The configuration of colour-charges in ${\sf d}^\circ_2$ allows 
a third doublet (a pair of unlike-charged tripoles) to be attached 
to the ends of the chain:
\begin{equation}
{\sf d}^\circ_{3}= 
\DLN \hspace{1.5ex} \DUN \hspace{0.6ex} {\DRN}\hspace{-0.4ex}\DoT{\DRP} 
\DoT{\DUP} \hspace{-0.2ex}\DoT{\DLP}~.
\label{eq:dzerothreeup}
\end{equation}
This completes all the three possible $\frac{2}{3}\pi$-rotations
of tripoles in the chain. 
The position angle between the 
tripoles at the loose ends of the chain (\ref{eq:dzerothreeup}) 
is $180^\circ$, which 
corresponds to the attractive force between these tripoles
[see the diagram (\ref{eq:dipoleattractive})]. This allows
closing the chain into a loop:
\begin{equation}
\DLN
\hspace{-0.1cm}
\raisebox{-0.02cm}
   {$
     \stackrel{
	       \raisebox{0.4cm}
                 {
                  {\begin{turn}{15}{$\diagup$}\end{turn}}
                 }
               }
              {
               \raisebox{-0.4cm}
                 {
                  {\begin{turn}{-15}{$\diagdown$}\end{turn}}
                 }
              }
   $}
\hspace{-0.3cm}
\raisebox{-0.05cm}
  {$
    \stackrel{
              \raisebox{0.9cm}
                 {
                  {\DUN}\hspace{0.5ex}---%
                  \hspace{0.2ex} {\DRN}
                 }
             }
             {
              \raisebox{-0.9cm}
                {
                 \hspace{-0.7ex}\DoT{\DLP}%
\hspace{0.2ex}---\hspace{0.2ex}\DoT{\DUP}
                }
             }
  $}
\hspace{-0.4cm}
\raisebox{-0.02cm}
   {$
     \stackrel{
	       \raisebox{0.5cm}
                 {
                  {\begin{turn}{-15}{$\diagdown$}\end{turn}}
                 }
               }
              {
               \raisebox{-0.5cm}
                 {
                  {\begin{turn}{15}{$\diagup$}\end{turn}}
                 }
              }
   $}
\hspace{-0.2cm}
\raisebox{-0.05cm}
  {
   \hspace{-0.7ex}\DoT{\DRP}~,
  }
\label{eq:gammanurightcycle}
\end{equation}
the constituents of which can recombine then into a string of three
neutral doublets:
\begin{equation}
{\sf d}^\circ_3 \rightarrow 3{\sf d}^\circ= 
\hspace{0.4ex} \DLN \hspace{0.2ex}\DoT{\DLP} \hspace{0.6ex}  
\DUN \hspace{-0.4ex} \DoT{\DUP} \hspace{0.6ex} 
{\DRN} \hspace{-0.4ex}\DoT{\DRP}~.
\label{eq:threedzero}
\end{equation}
This restructuring would happen because of the mutual attraction 
between the unlike-charged tripoles in the loop
(see Fig.\ref{fig:tantit}: 
the position angle $\zeta=180^\circ$ between unlike-charged 
tripoles corresponds to a potential well).
The string $3{\sf d}^\circ$ must be unstable, but a longer chain of 
the tripole-antitripole pairs possesses a kind of cyclic symmetry that 
allows its closure into a symmetric stable ring.
Indeed, the potential well extending from $\zeta=120^\circ$
to $\zeta=240^\circ$ (Fig.\ref{fig:tantit}) implies the possibility
of the unlike-charged tripoles in the chain being mutually rotated
by $120^\circ$. Depending on the chosen direction of rotation
(clockwise or anticlockwise), the chain can have 
one of two possible patterns of colour charges:  
\begin{equation}
{\sf d}_{6\up}^\circ= 
\DUN\hspace{0.6ex} {\DRP}\hspace{-0.7ex}\DoT{\DRP}~ \DoT{\DLN}{\DLN} 
\hspace{1.0ex}{\DUP}\DoT{\DUP}\hspace{-0.1ex} \DoT{\DRN} {\DRN}
 \hspace{0.6ex} \DLP \DoT{\DLP} \hspace{0.4ex}\DoT{\DUN}
\hspace{0.5mm}\dots
\label{eq:nuright}
\end{equation}
 or
\begin{equation}
{\sf d}_{6\down}^\circ=
\DUN ~ \DLP \DoT{\DLP} \DoT{\DRN} \DRN ~ {\DUP}\hspace{-0.5ex}\DoT{\DUP}
\DoT{\DLN}{\DLN}~~{\DRP}\hspace{-0.6ex}\DoT{\DRP} \DoT{\DUN} 
\hspace{0.5mm}\dots~.
\label{eq:nuleft}
\end{equation}
The pattern repeats after each six consecutive links (doublets). 
The mutual orientation of the like-charged tripoles in the first 
and sixth links (their $180^\circ$-position angle) corresponds 
to the attractive force between them and  allows the   
closure of ${\sf d}^\circ_6$ in a hexagonal loop 
with six tripole-antitripole pairs (hexaplet) 
\begin{equation}
\DoT{\DUN}\DUN
\hspace{-0.4cm}
\raisebox{-0.02cm}
   {$
     \stackrel{
	       \raisebox{0.4cm}
                 {
                  {\begin{turn}{15}{$\diagup$}\end{turn}}
                 }
               }
              {
               \raisebox{-0.4cm}
                 {
                  {\begin{turn}{-15}{$\diagdown$}\end{turn}}
                 }
              }
   $}
\hspace{-0.4cm}
\raisebox{-0.05cm}
  {$
    \stackrel{
              \raisebox{0.9cm}
                 {
                  {\DRP}\hspace{-0.7ex}\DoT{\DRP}---%
\DoT{\DLN}\hspace{-0.7ex}{\DLN}
                 }
             }
             {
              \raisebox{-0.9cm}
                {
                 \DLP\hspace{-0.7ex}\DoT{\DLP}%
---\DoT{\DRN}\hspace{-0.7ex}\DRN
                }
             }
  $}
\hspace{-0.4cm}
\raisebox{-0.02cm}
   {$
     \stackrel{
	       \raisebox{0.5cm}
                 {
                  {\begin{turn}{-15}{$\diagdown$}\end{turn}}
                 }
               }
              {
               \raisebox{-0.5cm}
                 {
                  {\begin{turn}{15}{$\diagup$}\end{turn}}
                 }
              }
   $}
\hspace{-0.6cm}
\raisebox{-0.05cm}
  {
   \DUP\hspace{-0.7ex}\DoT{\DUP}~
  }
\label{eq:nurightcycle}
\end{equation}
which we shall denote
\begin{equation*}  
{\tt X} \equiv {\sf d}^\circ_6=6\times{\left({~\DP~\DN}\right)}.
\end{equation*}
As in the case of the triplet ${\tt Y}$, the helical trajectory 
of any particular colour-current in the hexaplet is clockwise 
(${\tt X}_\up$) or anticlockwise (${\tt X}_\down$), 
which, by its closure,  makes a $\pi$-twist around the hexaplet's 
torus.
The total number of charges in ${\tt X}$ is 36, which corresponds to
twelve $\triangle$-shaped tripoles:
\begin{equation*} 
n_{\tt X}=12\times 3=36.
\end{equation*}
The hexaplet is neutral and almost massless, according to 
(\ref{eq:mass}). 
As in the case of the triplet ${\tt Y}$, 
the motion of charges in the hexaplet along 
its ring-axis is synchronised with their spin around this axis and with 
radial oscillations of the structure. 
Under an external electric field, the structure will be polarised.
In this case, the radial shifts of opposite charges 
$(~\DP\hspace{-0.8ex}\DoT{\DP}\uparrow  \downarrow ~ 
\DN \hspace{0.4ex}\DoT{\DN})$ can be in phase ($\phi=0$) or out of phase 
($\phi=\pi$) with respect to the fluctuations of $\rho_{\tt X}$.  One of these
phases corresponds to the positive charges~ $\DP \hspace{-0.8ex}\DoT{\DP}$
transferred to the outermost rim of the hexaplet's torus 
when $\rho_{\tt X}$ is maximal (and, respectively, 
the negative charges sitting on this rim when  
$\rho_{\tt X}$ is minimal): 
\begin{equation}
\hspace{1.0ex}\DoT{\DUN}{\DUN}
\hspace{-4.0ex}
\raisebox{-0.2ex}
   {$
     \stackrel{
	       \raisebox{4.0ex}
                 {
                  {\begin{turn}{-23}{$\diagdown$}\end{turn}}
                 }
               }
              {
               \raisebox{-4.0ex}
                 {
                  {\begin{turn}{23}{$\diagup$}\end{turn}}
                 }
              }
   $}
\hspace{-5.0ex}
\raisebox{-0.5ex}
  {\hspace{-0.5ex}$
    \stackrel{
              \raisebox{7.0ex}
                 {
		  \hspace{-0.5ex}
                  {\DRP}\hspace{-0.7ex}\DoT{\DRP}
	           \hspace{-1.5ex}
                   \raisebox{-1.0ex}
                    {
                      {\begin{turn}{5}$\diagdown$\end{turn}}
                    }
                   \raisebox{-4.0ex}
                     {
	              \hspace{-3.0ex}\DoT{\DLN}\hspace{-0.7ex}{\DLN}
                     }
                 }
             }
             {
              \raisebox{-7.0ex}
                {
                 \hspace{-0.5ex}
                 {\DLP}\hspace{-0.7ex}\DoT{\DLP}
                 \hspace{-1.5ex}
                 \raisebox{3.0ex}
                  {
                    {\begin{turn}{-5}$\diagup$\end{turn}}
                  }
                 \raisebox{4.0ex}
                   { 
                     \hspace{-3.0ex}\DoT{\DRN}\hspace{-0.7ex}{\DRN}
                   }
                }
             }
  $}
\hspace{-2.0ex}
\raisebox{-1.0ex}
   {$
     \stackrel{
	       \raisebox{0.1ex}
                 {
                    \hspace{-1.0ex}
                    {\begin{turn}{35}{$\diagdown$}\end{turn}}
                 }
               }
              {
               \raisebox{-0.1ex}
                 {
                    \hspace{-1.0ex}
                  {\begin{turn}{-35}{$\diagup$}\end{turn}}
                 }
              }
   $}
\hspace{-1.5ex}
\raisebox{-0.5ex}
  {
   {\DUP}\hspace{-0.7ex}\DoT{\DUP}~
  }.
\label{eq:nuright2}
\end{equation}
Another phase corresponds to the negative charges~ 
$\DN \hspace{0.2ex} \DoT{\DN}$ on the outermost rim of the structure
at the moment of $\rho_{\tt X}^{\rm max}$:
\begin{equation}
\DoT{\DUN}\hspace{-0.1ex}{\DUN}
\hspace{-0.5ex}
\raisebox{-0.2ex}
   {$
     \stackrel{
	       \raisebox{-1.0ex}
                 {
                  {\begin{turn}{-35}{$\diagup$}\end{turn}}
                 }
               }
              {
               \raisebox{-2.0ex}
                 {
                  {\begin{turn}{30}{$\diagdown$}\end{turn}}
                 }
              }
   $}
\hspace{-1.25ex}
\raisebox{-0.5ex}
  {$
    \stackrel{
              \raisebox{3.5ex}
                 {
                  \hspace{-0.5ex}
                  \DRP\hspace{-0.6ex}\DoT{\DRP}
                  {\raisebox{3.0ex}
                    {
                      \hspace{-1.5ex}
                      \begin{turn}{-5}{$\diagup$}\end{turn}
                    }
                  }
                  {\raisebox{5.0ex}
                    {
                     \hspace{-3.5ex}
                      \DoT{\DLN}\hspace{-0.6ex}{\DLN}
                    }
                  } 
                 }
             }
             {
              \raisebox{-3.5ex}
                {
                 {\DLP}\hspace{-0.6ex}\DoT{\DLP}
                 {\raisebox{-3.0ex}
                   {
                     \hspace{-1.5ex}
                     \begin{turn}{7}{$\diagdown$}\end{turn}
                   }
                 }
                 {\raisebox{-5.0ex}
                   {
                   \hspace{-3.5ex}
                   \DoT{\DRN}\hspace{-0.6ex}{\DRN}
                   }
                 }
                }
             }
  $}
\hspace{-3.5ex}
\raisebox{-0.2ex}
   {$
     \stackrel{
	       \raisebox{4.0ex}
                 {
                  \hspace{-3.0ex}
                  {\begin{turn}{20}{$\diagup$}\end{turn}}
                 }
               }
              {
               \raisebox{-4.0ex}
                 {
                  \hspace{-3.0ex}
                  {\begin{turn}{-20}{$\diagdown$}\end{turn}}
                 }
              }
   $}
\hspace{-5.5ex}
\raisebox{-0.5ex}
  {
   \DUP\hspace{-0.6ex}\DoT{\DUP}~
  } \hspace{0.2cm}.
\label{eq:nuleft2}
\end{equation}
This implies the possibility of two dynamical 
polarisation modes,  ~${\tt X}^\pm$ and 
${\tt X}^\mp$ (the oscillating structure with these two
phases can be denoted as $\tilde{\tt X}$).
Of course, the net charge of the structure remains zero. 
However, this is not so of the magnetic field: due to the 
non-uniformity of the innermost and outermost 
parts of this field the hexaplet
will possess a non-vanishing residual magnetic field, with 
$\overrightarrow{B}_{\tt X}$
parallel to the hexaplet's vector of angular momentum 
$\overrightarrow{\ell}_{\tt X}$ in the case of the positive 
polarisation mode: 
\begin{equation}
\overset{\hspace{-2.0ex}\stackrel{{\tiny \ell}\rightarrow}
{\raisebox{0.1ex}{\tiny {\sf B}$\rightarrow$}}}{{\tt X}^\pm}
\label{eq:hmomentumplus}
\end{equation}
and antiparallel in the case of the negative polarisation:
\begin{equation}
\overset{\stackrel{\hspace{-2.0ex}{\tiny \ell}\rightarrow}
{\raisebox{0.1ex}{\tiny $\leftarrow$ \hspace{-0.9ex}{\sf B}}}}{{\tt X}^\mp}.
\label{eq:hmomentumminus}
\end{equation}

The ``instantaneous'' view of the hexaplet
(for one of its phases) is shown in Fig.\ref{fig:twonu}.
\begin{figure}[htb]
\centering 
\epsfysize=2cm
\includegraphics[scale=0.2]{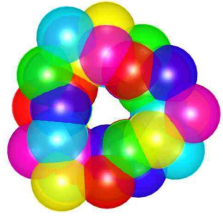}
\caption{
 Six-component string ${\tt X}$ (hexaplet) formed of the unlike-charged
 tripole pairs spinning around their common ring-closed axis. 
Unlike Fig.\ref{fig:tripletcurrents}, where each colour charge is
chown with its trajectory along the ring axis, the diagram above 
shows a ``snap-shot'' ( instantaneous configuration)
of the charges constituting the hexaplet.
       }
\label{fig:twonu}
\end{figure}
\section{Combinations of closed strings and tripoles}

The particles ${\tt Y}$,  ${\tt X}$, as well as the $\triangle$-shaped
tripoles,  are similar to each other and can combine  
because of short-range forces caused by the residual chromaticism
 (known as gluonic van der Waals forces) of these structures.
By considering spatial configurations of colour-charges
in a pair of ring-closed structures, say ${\tt XX}$, ${\tt YY}$,
or ${\tt XY}$, one can note that the sign of the 
van der Waals force depends on the particle helicities. 
Let us take a pair of  like-charged ${\tt Y}$-particles with 
opposite helicities:
\begin{equation}
{\tt Y}^+_\up=
\begin{bmatrix} \DoT{\DoT{\DLP}} \\ \DoT{\DoT{\DUP}} \\ 
\DoT{\DoT{\DRP}} \end{bmatrix}^\up
\hspace{2.0ex} \text{and} \hspace{3.0ex}
{\tt Y}^+_\down=
\begin{bmatrix} \DoT{\DLP} \\ \DoT{\DUP} \\ \DoT{\DRP} 
\end{bmatrix}_\down
\hspace{2.0ex}.
\label{eq:yplusyplus}
\end{equation}
Let us assume that the rotational and oscillatory frequencies
and phases of these particles are synchronised. In this regime
the mutual orientation of the particle constituents remains 
always the same, which would simplify the analysis 
of such a combined system.
The systems with no correlation between their
moving constituents are not solvable in principle. 
Then, a probabilistic approach
in the description of such systems would be more 
appropriate.

Note that in the representation (\ref{eq:yplusyplus}) 
we mark the first 
component of each structure with the symbols 
$\up$ and $\down$ (denoting the clockwise and anticlockwise 
rotations, respectively) -- to avoid possible confusion 
in the direction of rotation (these marks are not 
necessary in the case of a three-dimensional representation, 
as seen in Fig.\ref{fig:form3y}a). 

The $\triangle$-shaped tri\-poles -- the components of the 
structure ${\tt Y}^+_\up {\tt Y}^+_\down$ --
 are grouped in pairs and pair-wisely rotated by 
the position angle $\zeta=\pi$, which corresponds to the 
attractive force between them [see Fig.\ref{fig:yydoublet}(b) 
for $\rho>2\rho_\circ$]. Thus, besides the usual repulsive force 
between like-charges, the structure (\ref{eq:yplusyplus}) has 
an additional -- attractive --
force between its componets:   
\begin{equation}
\overset{\hspace{-1.0ex}\Leftarrow}
{
\begin{bmatrix} \DoT{\DoT{\DLP}} \\ \DoT{\DoT{\DUP}} \\ 
\DoT{\DoT{\DRP}} \end{bmatrix}^\up
}
\hspace{-1.0ex}
\begin{matrix} & \rightarrow & \leftarrow \\
               & \rightarrow & \leftarrow \\
               & \rightarrow & \leftarrow
\end{matrix} 
\overset{\Rightarrow}
{
\begin{bmatrix} \DoT{\DLP} \\ \DoT{\DUP} \\ \DoT{\DRP} 
\end{bmatrix}_\down
} ~.
\label{eq:yyupdown}
\end{equation}
This implies the possibility of an equilibrium state (merger) of
the pair, provided that there are no external forces and that the 
relative momenta of ${\tt Y}^+_\up$ and ${\tt Y}^+_\down$ 
are small enough. In this (entangled) state the orientation of the two 
particles is mutually dependent, which 
corresponds to the local minumum of the combined effective 
potential of the structure. 

In the case of like-helicities, two of the three triplet pairs 
have the position angle $\zeta=\pi/6$ between the constititing tripoles, 
which would cause their repulsion from each other: 
\begin{equation}
\begin{matrix}   ~ \\
                \leftarrow \\
                \leftarrow
\end{matrix} 
\overset{\hspace{-1.0ex}\Leftarrow}
{
\begin{bmatrix} \DoT{\DoT{\DLP}} \\ \DoT{\DoT{\DUP}} \\ 
\DoT{\DoT{\DRP}} \end{bmatrix}^\up
}
\hspace{-1.0ex}
\begin{matrix} & \rightarrow & \leftarrow \\
               & ~ & ~ \\
               & ~ & ~
\end{matrix} 
\overset{\Rightarrow}
{
\begin{bmatrix} \DoT{\DLP} \\ \DoT{\DRP} \\ \DoT{\DUP} 
\end{bmatrix}_\up
} 
\hspace{-2.0ex}
\begin{matrix} ~ \\
               & \rightarrow  \\
               & \rightarrow 
\end{matrix} 
\hspace{3.0ex}.
\label{eq:yyupup}
\end{equation}
This inhibits the entanglement of the particles 
with like-helicities.
It is noteworthy that the pattern of attraction and repulsion 
in the diagrams (\ref{eq:yyupdown})-(\ref{eq:yyupup})
coheres with (and probably explains the origin of) 
the Pauli exclusion principle.
%

In the merger (\ref{eq:yyupdown}) 
the {\tt Y}-particles are joined co-axially (pole-to-pole)
to each other. But they could also be coupled side-by-side
(laterally) provided that their vectors of angular momenta
are aligned, for instance, because of some external magnetic 
field.
In this case, the pattern of attraction for
the opposite helicities and repulsion for the like-helicities 
will also be reproduced. This means that every colour charge would tend
to occupy a position always in front of two other colour-charges,
complementary to the first one.
The structure with the laterally coupled {\tt Y}-particles
can grow (in principle, indefinitely) as a ``two-dimensional'' 
hexagonal lattice, as shown in Fig.\ref{fig:hexlattice}.     
\begin{figure}[htb]
\centering
\epsfysize=2cm
\includegraphics[scale=0.30]{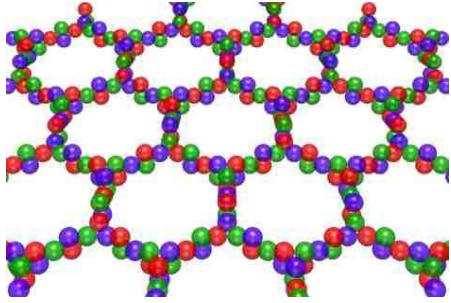}
\caption{A ``snap-shot'' of the hexagonal lattice formed 
by the side-by-side (off-axial) coupling of the triplets ${\tt Y}$.
} 
 \label{fig:hexlattice}
\end{figure}
Note that the lattice is shown here 
as a ``stand-still'' structure, although in reality its constituents 
are spinning (see Fig.\ref{fig:tripletcurrents}).
The stand-still representation 
is possible because of the above mentioned synchronisation of 
frequencies and phases between neighbouring {\tt Y}-particles. 

The two-dimensional infinite lattice, Fig.\ref{fig:hexlattice},
can hardly be stable. But it can be closed into more stable
configurations with lower energies, such as
a hollow tube (Fig.\ref{fig:hextorus}a), which
can further be closed in a torus (Fig.\ref{fig:hextorus}b).
The latter would be stable because its further closure
is unlikely (as with any other ring-structure).
The minimal torus can be   
formed of 108 ${\tt Y}$-triplets (${\tt T}_{108}$) or, alternatively, 
of 54 pairs ${\tt Y}^+{\tt Y}^-={\tt Y}^\circ$. In the latter case 
the particle ${\tt T}_{108}$ will be electrically neutral.
\begin{figure}[htb]
\centering
\epsfig{figure=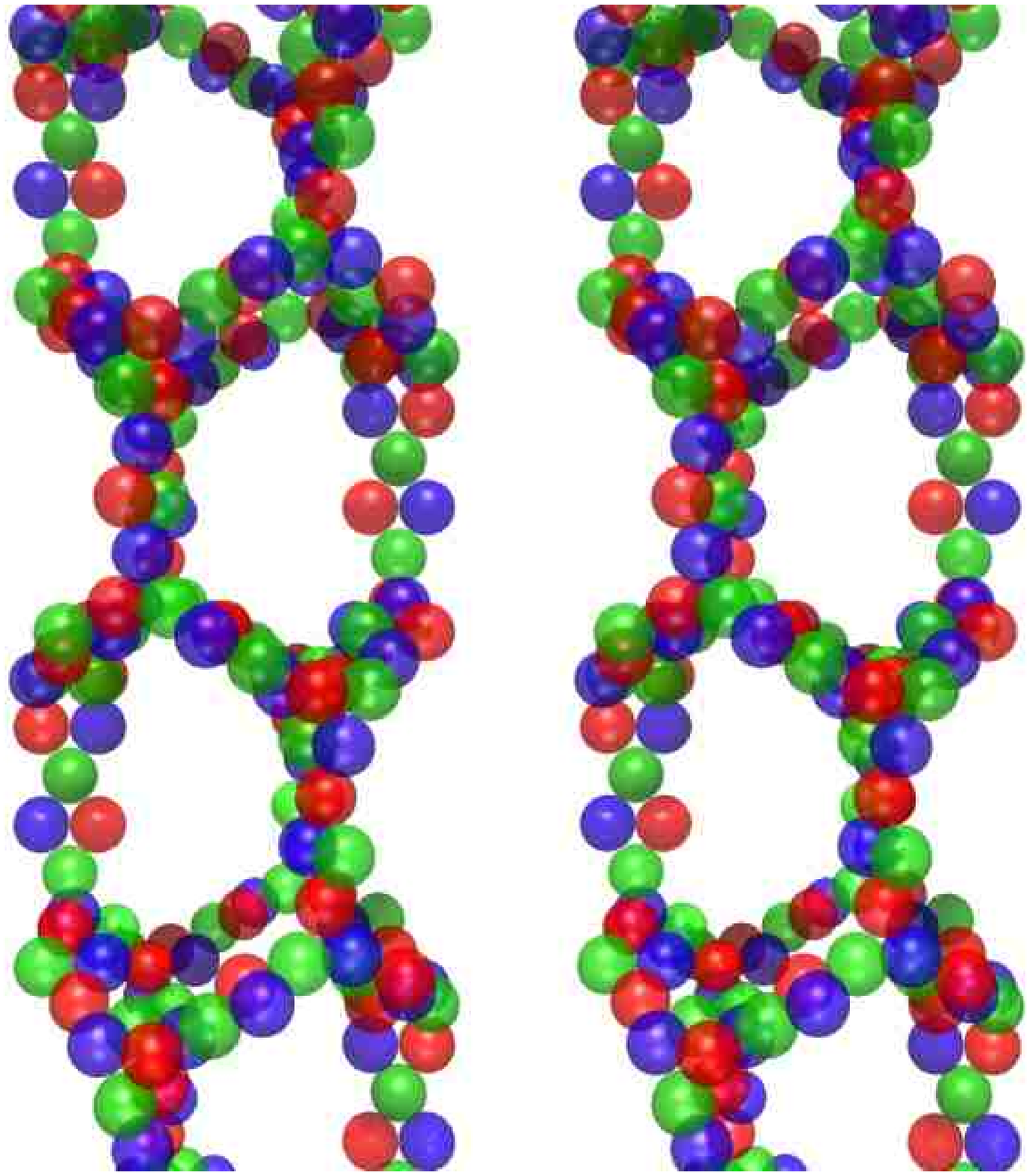,width=4.5cm}
\hspace{0.1cm}
\epsfig{figure=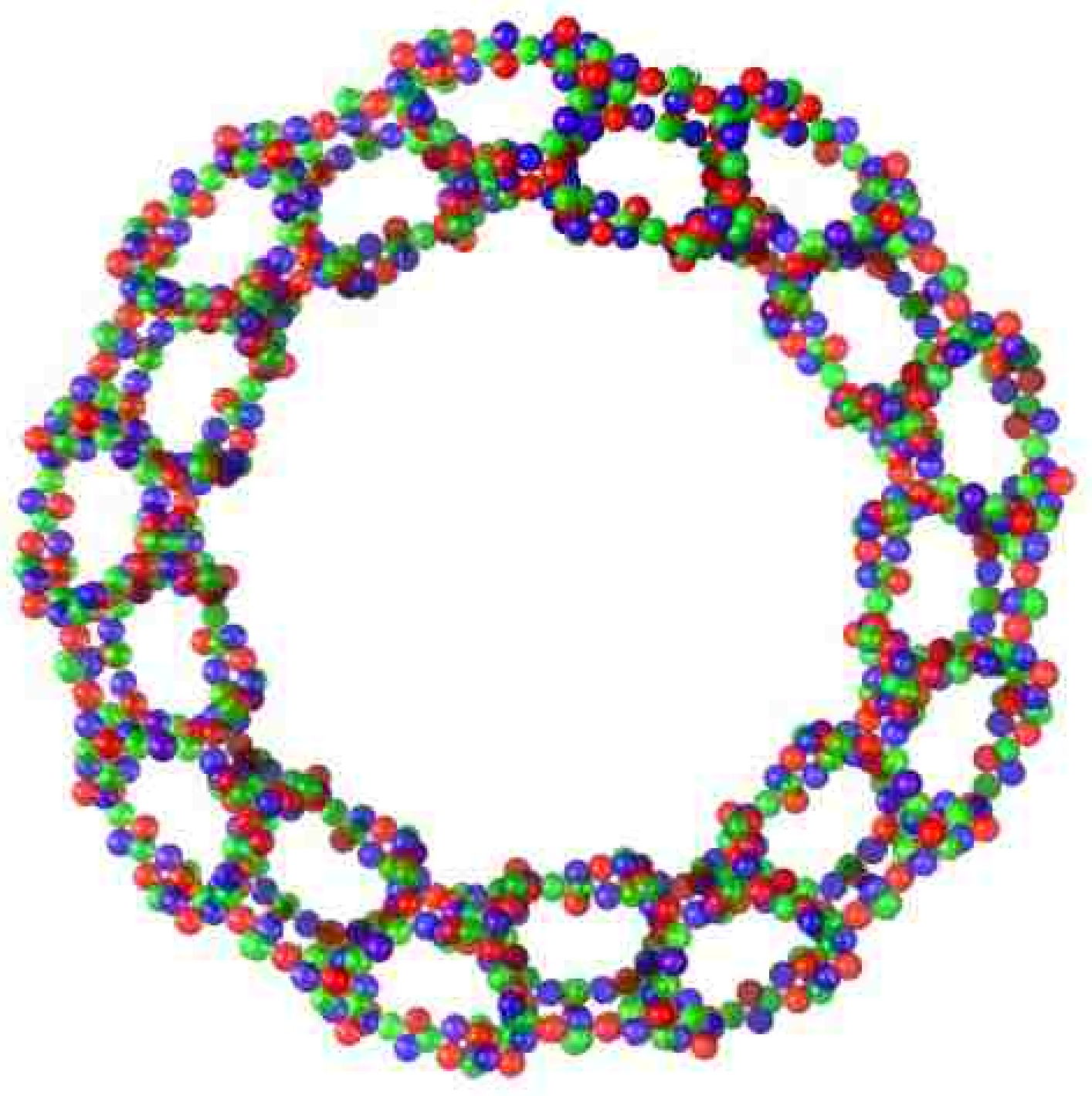,width=7.cm}
 \put(-275,2){
 \makebox(0,0)[t]{\footnotesize{(a)}}}
 \put(-110,2){
 \makebox(0,0)[t]{\footnotesize{(b)}}}
\caption{
(a) Hexagonal lattice of ${\tt Y}$-particles closed 
in a hollow tube; (b) the tube is further closed in a 
minimal-energy torus ${\tt T}_{108}$ of 108 ${\tt Y}$-particles 
(54 neutral pairs ${\tt Y}^+{\tt Y}^-$).
} 
 \label{fig:hextorus}
\end{figure}
Another possibility to form a stable neutral structure
of the triplets ${\tt Y}$ is the spherical closure of the 
lattice, as shown in Fig.\ref{fig:hexsphere}. 
The minimal number of ${\tt Y}$-particles that can combine in 
a hollow sphere is eight (or, respectively, four pairs of 
the oppositely charged ${\tt Y}^+$ and ${\tt Y}^-$ particles). 
\begin{figure}[htb]
\centering
\epsfysize=2cm
\includegraphics[scale=0.25]{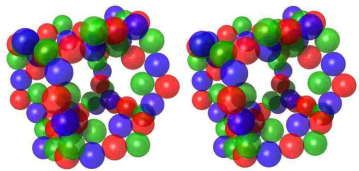}
\caption{Spherically-closed structure ${\tt S}_8$ composed of eight 
 triplets {\tt Y} (four neutral pairs ${\tt Y}^+{\tt Y}^-$).
} 
 \label{fig:hexsphere}
\end{figure}
%


Analysing the structural diagrams for the hexaplets ${\tt X}$ 
one will find that the dispersive  (van der Vaals)
force between ${\tt X}_\down$ and
 ${\tt X}_\up$ is attractive and between 
${\tt X}_\down$ and ${\tt X}_\down$ (or
${\tt X}_\up$ and ${\tt X}_\up$) -- repulsive. However,
in the case of a hexaplet combined with a triplet (${\tt XY}$)
the pattern of attraction and repulsion is inversed: the local 
chromaticism of ${\tt X}$ would be affine (attractive) 
to that of ${\tt Y}$ only if both particles have like-helicities.  

For the pairs of unlike-charged particles, ${\tt Y}^+$ 
and ${\tt Y}^-$, the pattern of attraction and repulsion 
is similar to (\ref{eq:yyupdown})-(\ref{eq:yyupup}), with 
the only difference that, in addition to this pattern, 
there exists the conventional attractive force between
opposite charges:
\begin{equation}
{\tt Y}^+_\up + {\tt Y}^-_\down = \hspace{1.0ex}
\overset{\hspace{-1.0ex}\Rightarrow}
{
\begin{bmatrix} \DoT{\DoT{\DLP}} \\ \DoT{\DoT{\DUP}} \\ 
\DoT{\DoT{\DRP}} \end{bmatrix}^\up
}
\hspace{-1.0ex}
\begin{matrix} & \rightarrow & \leftarrow \\
               & \rightarrow & \leftarrow \\
               & \rightarrow & \leftarrow
\end{matrix} 
\overset{\Leftarrow}
{
\begin{bmatrix} \DoT{\DLN} \\ \DoT{\DUN} \\ \DoT{\DRN} 
\end{bmatrix}_\down
} ~
\label{eq:yantiyupdown}
\end{equation}
\begin{equation}
{\tt Y}^+_\up + {\tt Y}^-_\up = \hspace{1.0ex}
\begin{matrix}   ~ \\
                \leftarrow \\
                \leftarrow
\end{matrix} 
\overset{\hspace{-1.0ex}\Rightarrow}
{
\begin{bmatrix} \DoT{\DoT{\DLP}} \\ \DoT{\DoT{\DUP}} \\ 
\DoT{\DoT{\DRP}} \end{bmatrix}^\up
}
\hspace{-1.0ex}
\begin{matrix} & \rightarrow & \leftarrow \\
               & ~ & ~ \\
               & ~ & ~
\end{matrix} 
\overset{\Leftarrow}
{
\begin{bmatrix} \DoT{\DLN} \\ \DoT{\DRN} \\ \DoT{\DUN} 
\end{bmatrix}_\up
} 
\hspace{-2.0ex}
\begin{matrix} ~ \\
               & \rightarrow  \\
               & \rightarrow 
\end{matrix} ~.
\label{eq:yantiyupup}
\end{equation}
The potential of the system (\ref{eq:yantiyupup}) has a 
repulsive inner and attractive outer region, whereas 
both the inner and outer regions of
the potential of the system (\ref{eq:yantiyupdown}) 
are attractive.
As a result, two unlike-charged ${\tt Y}$-particles would tend 
to merge, reaching an equilibrium (the neutral merger ${\tt Y}^\circ$ ) 
state between the constituents of both particles, which
 is possible if we suppose that the dynamics 
of the system is suppressed (say, by an external magnetic field).
Actually, this state:
\begin{equation}
{\tt Y}^\circ = \hspace{1.0ex}
\begin{bmatrix} \DoT{\DoT{\DLP}} \DoT{\DLN} \\ 
\DoT{\DoT{\DUP}} \DoT{\DUN} \\ 
\DoT{\DoT{\DRP}} \DoT{\DRN} \end{bmatrix}
 = \begin{bmatrix} {\sf d}^\circ \\ {\sf d}^\circ \\ {\sf d}^\circ 
 \end{bmatrix} ~
\label{eq:yyentangled}
\end{equation}
consists of three neutral doublets 
${\sf d}^\circ$ (\ref{eq:gamma}) 
joined together by their residual chromofields.
The ground-state energy (mass) of this neutral system will correspond to the 
masses of its two constituents, ${\tt Y}^+$ and ${\tt Y}^-$,
less the binding energy between them, $\hat{E}_{{\tt Y}^+{\tt Y}^-}$:
\begin{equation}
m_{{\tt Y}^\circ}\approx 2m_{\tt Y}-\hat{E}_{{\tt Y}^+{\tt Y}^-} 
< 18~[m_\circ]. 
\label{eq:mentangled}
\end{equation}
The bonds between these three doublets 
${\sf d}^\circ$ must be weak because there is no attractive 
electric force between neutral particles, and the doublets 
here are joined 
only by their residual chromofields. Thus, even 
small perturbations would cause this structure to desintegrate
into three neutral doublets.
The corresponding reaction can be written as
\begin{equation}
{\tt Y}^+_\up + {\tt Y}^-_\down = \hspace{1.0ex}
\overset{\hspace{-1.0ex}\Rightarrow}
{
\begin{bmatrix} \DoT{\DoT{\DLP}} \\ \DoT{\DoT{\DUP}} \\ 
\DoT{\DoT{\DRP}} \end{bmatrix}^\up
}
\hspace{-1.0ex}
\begin{matrix} & \rightarrow & \leftarrow \\
               & \rightarrow & \leftarrow \\
               & \rightarrow & \leftarrow
\end{matrix} 
\overset{\Leftarrow}
{
\begin{bmatrix} \DoT{\DLN} \\ \DoT{\DUN} \\ \DoT{\DRN} 
\end{bmatrix}_\down
} ~ \longrightarrow ~
\begin{matrix}
[\DoT{\DoT{\DLP}}\DoT{\DLN}] = {\sf d}^\circ \\
[\DoT{\DoT{\DUP}}\DoT{\DUN}] = {\sf d}^\circ \\
[\DoT{\DoT{\DRP}}\DoT{\DRN}] = {\sf d}^\circ
\end{matrix}
\label{eq:yantiyupdown2}
\end{equation}
or, in brief,
\begin{equation}
{\tt Y}^+_\up + {\tt Y}^-_\down 
\overset{t_{\up \down}}{\longrightarrow} 3{\sf d}^\circ,
\label{eq:eeupdown}
\end{equation}
where $t_{\up \down}$ is the time, which is needed to complete 
the reaction. Alternatively, the particles ${\tt Y}^+_\up$ 
and ${\tt Y}^-_\down$ can separate without loosing their
integrity, which would correspond to their elastic scattering.

In the system (\ref{eq:yantiyupup}) of a pair of  ${\tt Y}$-particles
with like-helicities, two of three tripole-antitripole
pairs have the position angles $\zeta=\pi/6$ corresponding 
to a repulsive force, which precludes direct merging of 
these components into neutral doublets. 
Instead, energetically, it is more economic for   
these two pairs of triplets to combine first into two charged 
doublets ${\sf d}^+$ and ${\sf d}^-$, Eq. (\ref{eq:deltapm}).
Only after that, these intermediate structures could merge and 
form a neutral system (\ref{eq:dzeroup}) ${\sf d}^\circ_2$:   
\begin{equation}
\raisebox{-1.5ex}
{
\DoT{\DoT{\DUP}}\DoT{\DoT{\DRP}}{\Big \{}
}
\hspace{-0.5ex}
\overset{\hspace{-1.0ex}\Rightarrow}
{
\begin{bmatrix} \DoT{\DoT{\DLP}} \\ \DoT{\DoT{\DUP}} \\ 
\DoT{\DoT{\DRP}} \end{bmatrix}^\up
}
\hspace{-1.0ex}
\raisebox{-1.0ex}
{$
\begin{matrix} & \hspace{0.5ex}\rightarrow & \hspace{-2.0ex} \leftarrow \\
               & ~ &  
\hspace{-4.5ex}
\raisebox{-2.0ex}
{
\DoT{\DRN}\DoT{\DUN}{\Big\{}
} \\
               & ~ & ~
\end{matrix} 
$}
\hspace{-1.0ex}
\overset{\Leftarrow}
{
\begin{bmatrix} \DoT{\DLN} \\ \DoT{\DRN} \\ \DoT{\DUN} 
\end{bmatrix}_\up
} 
\hspace{-2.0ex}
\hspace{2.0ex}
~\raisebox{0.5ex}
{$\longrightarrow$}
\raisebox{-1.0ex}
{$ 
\begin{matrix}
\hspace{2.0ex}
[\DoT{\DoT{\DLP}}\DoT{\DLN}] = {\sf d}^\circ \\
 ~ \\
\raisebox{3.0ex}
{
\hspace{-0.2ex}
[\DoT{\DoT{\DUP}}\DoT{\DoT{\DRP}}
\DoT{\DRN}\DoT{\DUN}]
 = {\sf d}$^\circ_2$
}
\end{matrix}
$}
\label{eq:yantiyupup2}
\end{equation}
with the energy of ${\sf d}^\circ_2$, twice as much as that
of ${\sf d}^\circ$. Since by its other properties
the particle  
${\sf d}_2^\circ$ is similar to  ${\sf d}^\circ$, 
we can write ${\sf d}^\circ_2 \sim {\sf d}^\circ$ and 
\begin{equation}
{\tt Y}^+_\up + {\tt Y}^-_\up \overset{t_{\up \up}}{\longrightarrow} 
2{\sf d}^\circ.
\label{eq:eeupup}
\end{equation}
The time  $t_{\up \down}$ in (\ref{eq:eeupdown}) will 
be smaller than  $t_{\up \up}$ in (\ref{eq:eeupup}) 
 because in the former case the 
inner part of the potential is attractive,  
whereas in the latter it is repulsive. 

Due to the replicated patterns and geometrical resemblance
between the tri\-poles $\triangle$, 
tri\-plets ${\tt Y}$, and hexa\-plets ${\tt X}$ (all possess the 
$\frac{2}{3}\pi$-symmetry of their constituent colour charges), 
it is not difficult to deduce how these structures can 
combine with each other. Obviously, the hexaplet ${\tt X}$, formed 
of twelve $\triangle$-tripoles, is geometrically larger than a 
single $\triangle$-tripole, 
thus, these two structures can combine only when the former 
enfolds the latter: 
\begin{equation}
\DoT{\DUN}\hspace{-0.1ex}{\DUN}
\hspace{-0.5ex}
\raisebox{-0.2ex}
   {$
     \stackrel{
	       \raisebox{-1.0ex}
                 {
                  {\begin{turn}{-35}{$\diagup$}\end{turn}}
                 }
               }
              {
               \raisebox{-2.0ex}
                 {
                  {\begin{turn}{30}{$\diagdown$}\end{turn}}
                 }
              }
   $}
\hspace{-1.25ex}
\raisebox{-0.5ex}
  {$
    \stackrel{
              \raisebox{3.5ex}
                 {
                  \hspace{-0.5ex}
                  \DRP\hspace{-0.6ex}\DoT{\DRP}
                  {\raisebox{3.0ex}
                    {
                      \hspace{-1.5ex}
                      \begin{turn}{-5}{$\diagup$}\end{turn}
                    }
                  }
                  {\raisebox{5.0ex}
                    {
                     \hspace{-3.5ex}
                      \DoT{\DLN}\hspace{-0.6ex}{\DLN}
                    }
                  } 
                 }
             }
             {
              \raisebox{-3.5ex}
                {
                 {\DLP}\hspace{-0.6ex}\DoT{\DLP}
                 {\raisebox{-3.0ex}
                   {
                     \hspace{-1.5ex}
                     \begin{turn}{7}{$\diagdown$}\end{turn}
                   }
                 }
                 {\raisebox{-5.0ex}
                   {
                   \hspace{-3.5ex}
                   \DoT{\DRN}\hspace{-0.6ex}{\DRN}
                   }
                 }
                }
             }
  $}
\hspace{-3.5ex}
\raisebox{-0.2ex}
   {$
     \stackrel{
	       \raisebox{4.0ex}
                 {
                  \hspace{-3.0ex}
                  {\begin{turn}{20}{$\diagup$}\end{turn}}
                 }
               }
              {
               \raisebox{-4.0ex}
                 {
                  \hspace{-3.0ex}
                  {\begin{turn}{-20}{$\diagdown$}\end{turn}}
                 }
              }
   $}
\hspace{-5.5ex}
\raisebox{-0.5ex}
  {
   \DUP\hspace{-0.6ex}\DoT{\DUP}~
  }
\hspace{-11.0ex}
\raisebox{-0.5ex}
 {
  $\overset{\begin{turn}{-58}{--}\end{turn}}{\underset
    {
      \raisebox{1.2ex}
        {
         \begin{turn}{58}{--}\end{turn}
        } 
    }{\hspace{2.0ex}\DLN}}$~--}
\label{eq:y1}
\end{equation}
Here the hexaplet would acquire a dynamical 
polarisation $(\mp)$ because of the 
presence in its centre of a negative charge. 
For brevity we shall denote this configuration
$\triangle^1$ (because of its resemblance with the simple 
tripole $\triangle$, albeit on a higher level of complexity):
\begin{equation} 
\DN^1=\left(\underset{\NN}{{\tt X}^\mp}\right).
\label{eq:delta1}
\end{equation}
In this notation the lower and upper parts 
of  the symbol ${\tt X}$ are used to denote, respectively, the 
innermost and outermost rims of the hexaplet's
torus. Then, placing \NN ~below ${\tt X}$
denotes the attachment of the triplet to the 
innermost rim of ${\tt X}$.
The structure $\triangle^1$ will be charged, with its charge   
$q_{{\triangle}^1}=- 3$ (or $+3$) derived from the simple  
$\triangle$-tripole, and will have a mass (since it is charged)
\begin{equation} 
m_{{\triangle}^1}=n_{\tt X}+m_{\triangle}=36+3=39 
\label{eq:massdelta1}
\end{equation}
units of $m_\circ$.
Similarly to the structure (\ref{eq:y1}), a single $\triangle$-shaped 
tripole can be enfolded with a triplet ${\tt Y}$:
\begin{equation}
\hspace{-13.2ex}
\underset{\NP}{\DoT{{\tt Y}}_-} =
\raisebox{2.5ex}
    {
     \begin{turn}{32}\DoT{\DLN}\end{turn}
    }
\hspace{-4.25ex}
\raisebox{-3.5ex}
   {
    \begin{turn}{-30.5}{\DRN}\end{turn}
                }
\hspace{1.5ex}
\raisebox{0.01ex}
  {
   \begin{turn}{-30.5}{\DLN}\end{turn}
  }
\hspace{-9.5ex}
\raisebox{-0.5ex}
 {
  $\overset{\begin{turn}{-58}{--}\end{turn}}{\underset
    {
      \raisebox{1.2ex}
        {
         \begin{turn}{58}{--}\end{turn}
        } 
    }{\hspace{2.0ex}\DLP}}$~--}
\hspace{2ex}
\label{eq:ydelta}
\end{equation}
(of course, if both of these particles are oppositely charged).
The structure (\ref{eq:ydelta}), like the structures 
(\ref{eq:y1}) and $\triangle$, cannot be free. However, it can 
form part of some more complex structures. The triplet ({\tt Y}) can
also enfold a neutral doublet (${\sf d}^\circ=\DN \NP$):
\begin{equation}
\underset{\DN \NP}{\DoT{{\tt Y}}_-}~.
\label{eq:triplet0doublet}
\end{equation}
Since both {\tt Y} and ${\sf d}^\circ$ have their diverging potentials 
closed/cancelled, the combination (\ref{eq:triplet0doublet}) can, 
in principle, be found in free states.

\section{Oscillating structure {\tt XY}}

The hexaplet ${\tt X}$ must be stiffer than the triplet ${\tt Y}$ 
because of stronger bonds between the unlike-charged components
of the former, while the repulsion between the like-charged
components of the latter makes the bonds between these components 
weaker. Then, the amplitude of possible oscillations
of $\rho_{\tt Y}$ is larger than 
that of $\rho_{\tt X}$. Thus, in the structure ${\tt XY}$,
it is the triplet that would enfold the hexaplet:
\begin{equation}
\hspace{1.0ex}\DoT{\DUN}{\DUN}
\hspace{-11.0ex}
\raisebox{-0.6ex}
   {$
     \stackrel{
	       \raisebox{12.0ex}
                 {
                  \hspace{0.1ex}
                  {\DLN} \hspace{-2.0ex}
                  \raisebox{-2.0ex}
                   {
                    $\diagdown$
                   }
                 }
               }
              {
               \raisebox{-12.0ex}
                 {
                  \hspace{0.1ex}
                  {\DRN} \hspace{-2.0ex}
                  \raisebox{2.0ex}
                    {
                     {\begin{turn}{8}{$\diagup$}\end{turn}}
                    }
                 }
              }
   $}
\hspace{-2.0ex}
\raisebox{-0.2ex}
   {$
     \stackrel{
	       \raisebox{4.0ex}
                 {
                  {\begin{turn}{-23}{$\diagdown$}\end{turn}}
                 }
               }
              {
               \raisebox{-4.0ex}
                 {
                  {\begin{turn}{23}{$\diagup$}\end{turn}}
                 }
              }
   $}
\hspace{-5.0ex}
\raisebox{-0.5ex}
  {\hspace{-0.5ex}$
    \stackrel{
              \raisebox{7.0ex}
                 {
		  \hspace{-0.5ex}
                  {\DRP}\hspace{-0.7ex}\DoT{\DRP}
	           \hspace{-1.5ex}
                   \raisebox{-1.0ex}
                    {
                      {\begin{turn}{5}$\diagdown$\end{turn}}
                    }
                   \raisebox{-4.0ex}
                     {
	              \hspace{-3.0ex}\DoT{\DLN}\hspace{-0.7ex}{\DLN}
                     }
                 }
             }
             {
              \raisebox{-7.0ex}
                {
                 \hspace{-0.5ex}
                 {\DLP}\hspace{-0.7ex}\DoT{\DLP}
                 \hspace{-1.5ex}
                 \raisebox{3.0ex}
                  {
                    {\begin{turn}{-5}$\diagup$\end{turn}}
                  }
                 \raisebox{4.0ex}
                   { 
                     \hspace{-3.0ex}\DoT{\DRN}\hspace{-0.7ex}{\DRN}
                   }
                }
             }
  $}
\hspace{-2.0ex}
\raisebox{-1.0ex}
   {$
     \stackrel{
	       \raisebox{0.1ex}
                 {
                    \hspace{-1.0ex}
                    {\begin{turn}{35}{$\diagdown$}\end{turn}}
                 }
               }
              {
               \raisebox{-0.1ex}
                 {
                    \hspace{-1.0ex}
                  {\begin{turn}{-35}{$\diagup$}\end{turn}}
                 }
              }
   $}
\hspace{-1.5ex}
\raisebox{-0.5ex}
  {
   {\DUP}\hspace{-0.7ex}\DoT{\DUP}~
  }
\hspace{-1.5ex}
\raisebox{-0.6ex}
  {
   {---\hspace{0.7ex}\DUN}
  }
\label{eq:w1}
\end{equation}
%
%
%
%
causing the positive $(\pm)$ dynamical polarisation of ${\tt X}$.
We can denote this structure as $^-\hspace{-0.5ex}{\tt Y}_1^\pm$ or
\begin{equation*}
{\tt XY}=\overset{\DoT{${\tt Y}$}_-}
{{\tt X}^\pm}
\end{equation*}
where the ($\pm$)-superscript indicates that   
the hexaplet ${\tt X}$ is polarised here
due to the negatively charged component attached to 
its outermost rim. Obviously, the hexaplet is always polarized 
positively (${\tt X}^\pm$) when combined with
${\tt Y}^-$ and negatively (${\tt X}^\mp$) -- when combined
with ${\tt Y}^+$. The components of the system
${\tt XY}$ would oscillate along their common axis:
\begin{equation}
\begin{aligned}
{\tt X} \leftarrow & {\tt Y} \\
 & {\tt Y}  \rightarrow {\tt X}
\end{aligned} \hspace{2.0ex}.
\label{eq:oscillationsW}
\end{equation}
But, these oscillations can be suppressed by an external
field if, for example, both particles are confined within 
some other (more complicated) structure. 
Then the mass of the structure ${\tt XY}$ with the suppressed oscillations 
can be approximately estimated as being proportional to the number
of its constituents, that is,
\begin{equation}
m_{\tt XY}=n_{\tt X}+ m_{\tt Y}=36+9=45 ~[m_\circ]
\label{eq:massHY}
\end{equation}
(less the binding energy between {\tt X} and {\tt Y}).
The charge of ${\tt XY}$ would correspond to the nine-unit 
charge of the triplet ${\tt Y}$ (the charged constituent of the 
system), 
\begin{equation*}
q_{{\tt XY}}=q_{\tt Y}=-9 ~[q_\circ].
\end{equation*}
In the oscillatory regime the ground-state energy (mass) of ${\tt XY}$
will be roughly proportional to the frequency of its 
oscillations: 
\begin{equation}
\omega_{\tt XY} = 2\pi \sqrt{\frac{k}{\hat{m}_{\tt XY}}},
\label{eq:omegahy}
\end{equation}
where $k$ is the bond force constant between ${\tt X}$ and ${\tt Y}$, 
and $\hat{m}_{\tt XY}$ is the reduced mass of {\tt XY}:  
\begin{equation*}
\hat{m}_{\tt XY}^{-1}=m_{\tt X}^{-1}+m_{\tt Y}^{-1}.
\end{equation*}
Since $m_{\tt X}\approx 0$, the reduced mass is also vanishing,
which means that the frequency $\omega_{\tt XY}$ would be very high
(and so too be the mass of ${\tt XY}$).

In fact, the force between ${\tt X}$ and ${\tt Y}$ cannot be 
a linear function of the distance between the two entities, and,
strictly speaking, the oscillatory frequency of ${\tt XY}$ 
cannot be accurately represented by the 
formula (\ref{eq:omegahy}), which corresponds to the 
classical ideal oscillator.
The oscillations of ${\tt XY}$ would be disrupted when their amplitude 
reaches a point where the non-linear effects are dominant.
After that, the components of the structure ${\tt XY}$ 
will move independently of each other, in opposite 
directions along their common axis 
(the structure ${\tt XY}$ will cease to exist). 
Of course, the amplitude of the oscillations of ${\tt XY}$ 
cannot grow {\it per se}. But one can notice that 
this system is  asymmetric by its nature because of
the differences in the orientation of the magnetic 
fields and vectors of angular momenta   
of its two constituents. This would make 
the energy of one amplitude to grow at the expense of 
the other (of course, the net energy of the 
system is conserved). At some point the system would disrupt,
which must happen under the larger amplitude.   
Due to this, the decay products of two different  
systems, ${\tt Y}^+{\tt X}^\mp$ and ${\tt Y}^-{\tt X}^\pm$, 
will differ by the relative orientation of  
their vectors of angular and 
linear momenta. Indeed, during the oscillations of,
for example, the negatively charged structure 
${\tt Y}^-{\tt X}^\pm$, its constituents,
${\tt X}$ and ${\tt Y}$, twist back and forth along the 
lines of their common magnetic field, with ${\tt X}$ 
receiving a boost when the direction of its motion 
coincides with the direction of its residual magnetic 
field, and decelerating otherwise:
\begin{equation}
\begin{aligned}
 \raisebox{1.0ex}{\Big [}&\hspace{1.0ex}\overset{\stackrel{\hspace{-2.0ex}
{\tiny \ell}\rightarrow}
{\raisebox{0.1ex}{\tiny $\leftarrow$ \hspace{-0.9ex}{\sf B}}}}{{\tt Y}^-}
\raisebox{1.0ex}{\Big ]} 
{\longrightarrow}
\raisebox{1.0ex}{\Big [}\hspace{1.0ex}\overset{\hspace{-2.0ex}
\stackrel{{\tiny \ell}\rightarrow}
{\raisebox{0.1ex}{\tiny {\sf B}$\rightarrow$}}}{{\tt X}^\pm}
\raisebox{1.0ex}{\Big ]}
\underset{p}{\text{---}\hspace{-0.9ex}\longrightarrow}  \\
\underset{p}{\leftarrow} \hspace{1.0ex} 
\raisebox{1.0ex}{\Big [}\hspace{1.0ex}\overset{\hspace{-2.0ex}
\stackrel{{\tiny \ell}\rightarrow}
{\raisebox{0.1ex}{\tiny {\sf B}$\rightarrow$}}}{{\tt X}^\pm}
\raisebox{1.0ex}{\Big ]} \longleftarrow
 \raisebox{1.0ex}{\Big [}&\hspace{1.0ex}\overset{\stackrel{\hspace{-2.0ex}
{\tiny \ell}\rightarrow}
{\raisebox{0.1ex}{\tiny $\leftarrow$ \hspace{-0.9ex}{\sf B}}}}{{\tt Y}^-}
\raisebox{1.0ex}{\Big ]}
 ~~.
\end{aligned}
\label{eq:asymmetryminusa}
\end{equation}
Conversely, for the positively charged triplet and negatively 
polarised hexaplet the boost to ${\tt X}$ occurs when its  
$\overrightarrow{\ell}$ and $\overrightarrow{p}$ 
vectors are antiparallel:         
\begin{equation}
\begin{aligned}
&\raisebox{1.0ex}{\Big [}\hspace{1.0ex}\overset{\hspace{-2.0ex}
\stackrel{{\tiny \ell}\rightarrow}
{\raisebox{0.1ex}{\tiny {\sf B}$\rightarrow$}}}{{\tt Y}^+}
\raisebox{1.0ex}{\Big ]}
\longrightarrow 
\raisebox{1.0ex}{\Big [}\hspace{1.0ex}\overset{\stackrel{\hspace{-2.0ex}
{\tiny \ell}\rightarrow}
{\raisebox{0.1ex}{\tiny $\leftarrow$ \hspace{-0.9ex}{\sf B}}}}{{\tt X}^\mp}
\raisebox{1.0ex}{\Big ]}\underset{p}{\rightarrow} \\
\underset{p}{\longleftarrow}\hspace{-0.9ex}\text{---} \hspace{1.0ex}
\raisebox{1.0ex}{\Big [}\hspace{1.0ex}\overset{\stackrel{\hspace{-2.0ex}
{\tiny \ell}\rightarrow}
{\raisebox{0.1ex}{\tiny $\leftarrow$ \hspace{-0.9ex}{\sf B}}}}{{\tt X}^\mp}
\raisebox{1.0ex}{\Big ]} 
\longleftarrow
& \raisebox{1.0ex}{\Big [}\hspace{1.0ex}\overset{\hspace{-2.0ex}
\stackrel{{\tiny \ell}\rightarrow}
{\raisebox{0.1ex}{\tiny {\sf B}$\rightarrow$}}}{{\tt Y}^+}
\raisebox{1.0ex}{\Big ]}.
\end{aligned}
\label{eq:asymmetryplus}
\end{equation}
This leads to a higher probability that ${\tt X}^\pm$ will  exit
the system in its right-handed state
(when its vectors $\overrightarrow{\ell}$ and 
$\overrightarrow{p}$ are parallel):
\begin{equation*}
{\tt X}_R=
\hspace{1.0ex}\overset{\hspace{-2.0ex}
\stackrel{{\tiny \ell}\rightarrow}
{\raisebox{0.1ex}{\tiny {$p$}$\rightarrow$}}}{\hspace{-1.0ex}{\tt X}^\pm},
\end{equation*}
and  for ${\tt X}^\mp$} -- in its left-handed state:  
\begin{equation*}
{\tt X}_L=
\overset{\stackrel{\hspace{-2.0ex}
{\tiny \ell}\rightarrow}
{\raisebox{0.1ex}{\tiny $\leftarrow$ \hspace{-0.9ex}{$p$}}}}
{\hspace{-1.0ex}{\tt X}^\mp}.
\end{equation*}
The corresponding reactions can be written as:
\begin{equation}
{\tt X}^\pm {\tt Y}^- \rightarrow {\tt X}^\pm_R + {\tt Y}^-_L
\label{eq:preferableminus}
\end{equation}
and
\begin{equation}
{\tt X}^\mp {\tt Y}^+ \rightarrow {\tt X}^\mp_L + {\tt Y}^+_R.
\label{eq:preferableplus}
\end{equation}
Of course, in a reference frame moving faster than ${\tt Y}^-_L$ 
or ${\tt Y}^+_R$, these particles can be observed 
as ${\tt Y}^-_R$ and ${\tt Y}^+_L$, respectively. However, 
the particle ${\tt X}^\mp$ will always be observed as left-handed 
since it moves with the maximal possible speed because of 
its vanishing mass. 
Likewise, the particle ${\tt X}^\pm$ will have 
right-handed preference. This kind of symmetry is usually 
referred to as the conjugation of charge and parity (CP-symmetry).

\section{Hierarchy of structures}

Like the simple $\triangle$-tripoles, the ``enfolded'' ones, 
$\underset{\nabla}{{\tt X}}$, can 
combine with each other, forming doublets,
strings, ring-closed loops, etc.
However, there is a difference between $\triangle$ 
and $\underset{\hspace{0.3ex}\nabla}{{\tt X}}$: the latter 
possesses the helicity property (derived from its 
constituent hexaplet ${\tt X}$). When two  
unlike-charged particles $\underset{\hspace{0.3ex}\nabla}{{\tt X}}$
combine, their  polarisation modes and helicity signs are {\it always} 
opposite (simply because their central tripoles have  
opposite charges). These opposite helicities cause an additional 
attractive force between the two particles, 
as well as the usual attractive force corresponding to 
the opposite electric charges of \DN~ and \DP: 
\begin{equation}
\left(\underset{\NN}{{\tt X}^\mp}\right)
\underset{\rightarrow}{\Rightarrow} 
\underset{\leftarrow}{\Leftarrow} \left(\underset{\NP}{{\tt X}^\pm}\right).
\label{eq:numu0}
\end{equation}
%
This structure is similar to the merger (\ref{eq:yantiyupdown2}), which 
exists only for a short period of time (until  
it disintegrates into the neutral doublets ${\sf d}^\circ$).
However, the disintegration of the structure (\ref{eq:numu0}) can be 
prevented by an oscillating hexaplet, 
\begin{equation*}
\tilde{{\tt X}}={\tt X}^\pm \leftrightsquigarrow {\tt X}^\mp
\hspace{0.2ex},
\end{equation*} 
which would create a repulsive stabilising force between the particles 
$\underset{\NN}{{\tt X}^\mp}$ and $\underset{\NP}{{\tt X}^\pm}$: 
\begin{equation*}
\underset{\NN}{{\tt X}^\mp} \tilde{{\tt X}} \hspace{0.5ex}
\underset{\NP}{{\tt X}^\pm}~ =
\begin{cases}
\underset{\NN}{{\tt X}^\mp} {\tt X}^\mp \leftrightsquigarrow 
\underset{\NP}{{\tt X}^\pm}~ \\
\underset{\NN}{{\tt X}^\mp} \leftrightsquigarrow 
{\tt X}^\pm
\underset{\NP}{{\tt X}^\pm}~
\end{cases}
.
\end{equation*}
%
The sides of this structure can accommodate 
another pair of unlike-charged $\underset{\nabla}{\tt X}$-particles:
\begin{equation}
\underset{\NP}{{\tt X}^\pm} \tilde{{\tt X}}
\underset{\NN}{{\tt X}^\mp} \tilde{{\tt X}}
\underset{\NP}{{\tt X}^\pm} \tilde{{\tt X}}
\underset{\NN}{{\tt X}^\mp}\hspace{0.5ex},
\label{eq:nutau0}
\end{equation}
and so on, until the string becomes flexible enough to be able to  
close into a ring (thus precluding its further growth).
We can denote such a ring-closed (neutral) structure as 
\begin{equation}
{\tt X}_1=6\times~\underset{\NN}{{\tt X}^\mp} \tilde{{\tt X}} 
\underset{\NP}{{\tt X}^\pm}~,
\label{eq:h1}
\end{equation}
by analogy with the structure (\ref{eq:nurightcycle}): 
~${\tt X}=6\times~\DN~\DP~$ --
to indicate that both structures 
are alike.  The structure (\ref{eq:h1}) contains 468 primitive 
charges: 
\begin{equation}
6\times (39+39)=468
\label{eq:mass468}
\end{equation}
(we neglect the contribution from the neutral massless 
component).

As opposed to the case (\ref{eq:numu0}), two hexaplets, if both enfold 
like-charged triplets, will have like-helicity signs. 
The extra force between such hexaplets will {\it always} 
be repulsive (in addition to the usual repulsive force between two 
like-charges):
\begin{equation*}
\underset{\leftarrow}{\Leftarrow} 
\left(\underset{\NP}{{\tt X}^\pm}\right)\hspace{1.5ex}\vdots
\hspace{1.5ex} \left(\underset{\NP}{{\tt X}^\pm}\right)
\underset{\rightarrow}{\Rightarrow}
\hspace{0.2cm}.
\end{equation*}
Thus, two like-charged 
$\underset{\hspace{0.3ex}\nabla}{{\tt X}}$-particles would never
combine, unless there exists an intermediate 
hexaplet (${\tt X}^\mp$) 
between them, with the 
helicity sign opposite to that of the components of the pair
(negatively polarised in this case). This
could neutralise the repulsive force between the components and
allow the following combination:
\begin{equation}
\label{eq:uplink}
\stackrel{\raisebox{0.5cm}{\hspace{0.6cm} \rm \tiny Charge:}}
    {
     \stackrel{\raisebox{-0.1cm}{\hspace{0.7cm}$~~$}}
         {\raisebox{-0.75cm}{\hspace{-0.7cm}\rm \tiny Number of charges:}
         }
    }
\hspace{-0.4cm}
\raisebox{-0.3cm}{$\leftarrow$}
\hspace{-0.4cm}
\raisebox{-0.01cm}
 {$
                                     {\stackrel{\raisebox{0.6cm}{\hspace{0.6cm}\bf \tiny +3}}
                                          {
                                           \stackrel{{\raisebox{0.01cm}{~~}}}
                                               {
                                                \raisebox{-0.75cm}{\hspace{0.5cm}\bf \tiny 3}
                                               }
                                          }
                                     }
                                      \raisebox{-0.35cm}
                                      {\hspace{-0.3cm}$ 
                                      \stackrel{{\raisebox{-0.01cm}{\hspace{-0.3cm}$\left(\underset{\hspace{-0.1cm}\NP}{^\pm{\tt X}}\right)\Rightarrow$}}}
                                          {
                                            \raisebox{-0.4cm}{\hspace{-0.4cm}\bf \tiny 36}
                                          }
                                      \stackrel{{\raisebox{0.33cm}{${\tt X}^\mp $}}}
                                          {
                                             \raisebox{-0.4cm}{\hspace{-0.05cm}\tiny (36)}
                                          }
                                      \stackrel{{\raisebox{-0.01cm}{$\Leftarrow\left(\underset{\NP}{{\tt X}^\pm}\right)$}}}
                                          {
                                           \raisebox{-0.4cm}{\hspace{0.15cm}\bf \tiny 36}
                                          }
                                      \stackrel{\raisebox{0.65cm}{\hspace{-0.8cm}\bf \tiny +3}}
                                          { 
                                            \stackrel{\raisebox{0.33cm}{~}}
                                                {
                                                  \raisebox{-0.4cm}{\hspace{-0.8cm} \bf \tiny 3}
                                                }
                                          }
                                       $}
 $}
\stackrel{\raisebox{0.55cm}{\hspace{-0.1cm} \bf \tiny  =~+6}}
    {
     \stackrel{\raisebox{-0.1cm}{\hspace{0.7cm}$$}}
         {\raisebox{-0.73cm}{\hspace{-0.1cm}\bf \tiny  =~78}
         }
    }
\hspace{-1.1cm}
\raisebox{-0.3cm}{$\rightarrow$}  
\raisebox{0.01cm}{\hspace{-0.3cm}.}
\end{equation} 
The magnitude of its charge corresponds to the charge of 
two $\triangle$-tripoles; that is, $+6 [q_\circ]$. Its mass will be 
proportional to the number of the primitive charges constituting
its two charged components,
\begin{equation*} 
2\times (36+3)=78~[m_\circ].
\end{equation*}
In (\ref{eq:uplink}) the charges and masses of the charged components 
are indicated, respectively,  above and below the symbols corresponding 
to these components.  We neglect the contribution to this mass of the neutral 
(massless) component ${\tt X}^\mp$
[in (\ref{eq:uplink}) we have enclosed the number of its charges in 
parentheses].

The positively charged structure (\ref{eq:uplink})
can combine with the negatively charged structure,  
$^-{\tt Y}_1^\pm$, Eq.(\ref{eq:w1}), of 45-units mass: 
\begin{equation}
\label{eq:down3}
\stackrel{\raisebox{0.5cm}{\hspace{0.6cm} \rm \tiny Charge:}}
    {
     \stackrel{\raisebox{-0.1cm}{\hspace{0.7cm}$ $}}
         {\raisebox{-0.65cm}{\hspace{-0.7cm}\rm \tiny Number of charges:}
         }
    }
\raisebox{-1.0cm}
 {$ 
\stackrel{
          \underbrace{\stackrel{{\hspace{0.2cm} {\tiny \overline{\nu}_ee^-}}} 
                          {
                           \overbrace{\stackrel{{\raisebox{0.01cm}{$\overset{~\DoT{\tt Y}_-}{{\tt X}^\pm}$}}}
                                          {
                                           {\raisebox{-0.6cm}{\hspace{-0.3cm}\bf \tiny 36}}
                                          }
                                      \stackrel{\raisebox{0.8cm}{\hspace{-0.4cm} \bf \tiny -9}}
                                          {
                                              \stackrel{{\raisebox{0.01cm}{$ $}}}
                                                {
                                                 {\raisebox{-0.6cm}{\hspace{-0.3cm} \bf \tiny 9}}
                                                }
                                          }
                                     }
                          }
                      }
         }
    {
     \raisebox{-0.1cm}{\hspace{-0.1cm} \tiny 45}
    }
 $}
\hspace{0.2cm}
\raisebox{-1.1cm}
 {$
\stackrel{
          \underbrace{\stackrel{u}
                         { 
                          \overbrace{
                                     {\stackrel{\raisebox{0.5cm}{\hspace{0.3cm}\bf \tiny +3}}
                                          {
                                           \stackrel{{\raisebox{0.01cm}{~}}}
                                               {
                                                \raisebox{-0.95cm}{\hspace{0.3cm}\bf \tiny 3}
                                               }
                                          }
                                     }
                                      \raisebox{-0.35cm}
                                      {$
                                      \stackrel{{\raisebox{-0.01cm}{\hspace{-0.3cm}$\underset{\hspace{-0.1cm}\NP}{^\pm{\tt X}}$}}}
                                          {
                                            \raisebox{-0.6cm}{\hspace{-0.01cm}\bf \tiny 36}
                                          }
                                      \stackrel{{\raisebox{0.33cm}{${\tt X}^\mp $}}}
                                          {
                                             \raisebox{-0.6cm}{\hspace{0.03cm}\tiny (36)}
                                          }
                                      \stackrel{{\raisebox{-0.01cm}{$\underset{\NP}{{\tt X}^\pm}$}}}
                                          {
                                           \raisebox{-0.6cm}{\hspace{-0.1cm}\bf \tiny 36}
                                          }
                                      \stackrel{\raisebox{0.5cm}{\hspace{-0.4cm}\bf \tiny +3}}
                                          { 
                                            \stackrel{\raisebox{0.33cm}{~}}
                                                {
                                                  \raisebox{-0.6cm}{\hspace{-0.25cm} \bf \tiny 3}
                                                }
                                          }
                                       $}
                                   }
                        }
                   }
          }
     {
      \raisebox{-0.1cm}{\hspace{-0.1cm} \tiny 78}
     }
 $}
\stackrel{\raisebox{0.6cm}{\hspace{0.1cm} \bf \tiny  =~-3}}
    {
     \stackrel{\raisebox{-0.1cm}{\hspace{0.7cm}$$}}
         {\raisebox{-0.7cm}{\hspace{0.1cm}\bf \tiny  =~123}
         }
    }  
\raisebox{0.01cm}{\hspace{-1.0cm}.}
\end{equation} 
The resulting structure will have 
a mass of 123-units ($45+78=123~[m_\circ]$) 
and a charge of $+6-9=-3~[q_\circ]$.

Obviously, the hierarchy of the equilibrium configurations 
of colour charges can be continued. But, due to the complexity of 
the emerging structures we shall discuss them elsewhere. 


\section{Discussion} 

In author's view, the outlined scheme can be used for 
building a model of composite fundamental fermions.  
The structure {\tt Y}, by its properties, can be identified 
with the electron, the structure {\tt X} -- with the electron neutrino,
the structures (\ref{eq:uplink}) and (\ref{eq:down3}) --
with the first generation quarks. Of course, the 
mentioned structures are classical objects based on 
deterministic potentals, whereas the fundamental fermions 
are known to be quantum objects.
However, many people believe that there exists a deep 
structural continuity between classical and quantum mechanics 
that should be exploited 
\cite{dirac45, dirac63, mostafazadeh99, raju04}.
It is conceivable 
that quantum phenomena  could arise as 
a result of information loss due to non-reversible dissipative 
processes and self-organisation in deterministic systems 
when these systems undergo qualitative (phase) transitions
from one structural level to another
\cite{hooft99, hooft01, prigogine01}.
Thus, our conjecture is not at odds with the existing
quantum theories. 

The Standard Model of particle physics considers all 
the fundamental particles as point-like objects, whereas
there exists evidence of their compositeness.
The fact that the fundamental fermions fall into a nice 
pattern of three families suggests that there must exist   
some underlying structures that give rise to this pattern.
There is no obvious reason why there should 
be twelve fundamental particles with different properties. 
Considering these particles as the fundamental ones
is somewhat logically inconsistent. 
Most of them are unstable and decay into lighter 
fundamental particles. Then a reasonable question arises: 
How can the fundamental objects decay into 
equally fundamental ones?

Besides these logical reasons, there exists experimental 
(albeit still inconclusive) evidence of quark compositeness,
which  comes from proton-proton and positron-proton 
scattering experiments 
\cite{eichten83, adloff97, breitweg97}. These experiments
show that the probability of particle scattering for 
the most energetic collisions (that probe the distances below
1/1000th of the size of the proton or, equivalently, energies
above 200 GeV) is significantly higher than that predicted by
current theoretical models. 
The experiments with quark-quark scattering by the Collider 
Detector at Fermilab (CDF) group \cite{abe93, abe96} also 
showed evidence for substructure within the quark. 
Even though the later measurements made 
by the D0 collaboration \cite{abbott01} did not confirm 
the excess of the scattering probability for high  
energy jets, the results of all the scattering experiments,
taken in the context of the observed pattern of quark 
properties, make a strong point in favour of quark 
compositeness.

The $\frac{2}{3}\pi$-symmetrical structure of the electron 
revealed in this paper can also be tested by observations. 
For example, according to the discussed model, 
a monolayer of electrons at low temperatures should form 
a hexagonal lattice (similar to that shown in 
Fig.\ref{fig:hexlattice}). A laboratory set-up for 
such a test is feasible with current 
technology. The compositeness of the electron can 
also be seen by accurate measurements of the charges of its
three constituents. And indeed, the experimental evidence of 
fractional ($\frac{1}{3}e$) charges  was reported more than two 
decades ago \cite{larue81}.
High accuracy experiments aimed at detecting fractional 
charges were also conducted in 1997 in the Weizmann 
Institute of Science \cite{picciotto97} and in the 
CEA/Saclay laboratory \cite{saminadayar97}. 
Both groups measured a small electric current 
in a two-dimensional electron gas sandwiched between two 
semiconductor layers. The sample was cooled to less 
than 1K and a strong magnetic field was applied at right 
angles to the layers. By analysing the shot noise in 
this regime, both groups reported evidence that the
electric current is carried by quanta with charge 
one-third that of the electron. The conventional explanation 
of the fractional charges is based on quasiparticles 
(collective excitations in systems of many interacting
electrons) that exhibit fractional charges. 
However, quasiparticles are virtual objects. What was actually
observed in those experiments was the fractional charge itself 
but not quasiparticles. Our model explains  these facts 
in a much more straightforward way  
  -- by the compositeness of the electron.

It is obvious that if the fundamental fermions are 
bound states of smaller entities then by disregarding 
their possible structures one could never explain the 
origin of their properties, nor 
the spectrum of their masses, no matter how advanced 
and complex were the mathematical tools used. 
Taking an example from molecular physics: perhaps nobody 
would seriously propose to explain
the structure of, say, the carbon molecule by combining 
the mechanical, optical and electrical parameters of graphite
or diamond through symmetrical matrices. On the contrary,
the inverse procedure of analysing the variety of 
equilibrium configurations of carbon atoms yielded 
the prediction of carbon nanotubes and fullerenes 
\cite{ajayan99,harris99}. Firstly, most of the physicists 
denied the existence of these exotic
molecules. But at present, the molecules C$_{60}$, C$_{70}$, 
C$_{76/78}$ and C$_{84}$ are routinely supplied by 
commercial companies, and the immense industrial potential
of using carbon nanotubes is broadly realised as well. 

In our approach we use a similar inverse procedure 
by guessing at the basic symmetries 
of space and deriving the equilibrium particle 
configurations allowed by these symmetries. 
Each structure contains a well-defined number 
of constituents corresponding to the configuration with 
the lowest energy. So, the number of these constituents
[e.g., in the structures (\ref{eq:h1}), (\ref{eq:uplink}),
or (\ref{eq:down3})] 
is not a free parameter of the model but rather a fixed  
quantity determined by the basic symmetry of the potential. 
Likewise, the number of atoms in a crystal or a cyclic molecule
cannot be considered as a free parameter of a model describing
this crystal or molecule.

Of course, the idea of the fermion compositeness is not new.
Most of the existing composite models describe each quark and 
lepton as a combination of three sub-quark particles usually 
called ``pre-quarks'' or ``preons'' 
(see, e.g. \cite{pati73,harari79, harari81}). 
But at present the preon models are not very popular because they face
grave problems with gauge anomalies and divergences on small scales
\cite{weinberg76, dugne97, peccei97}.

The main problem is that of the preon's mass. 
It is known from scattering 
experiments that quarks and leptons are ``point-like'' down 
to distance scales of less than $10^{-18}$ m (or 1/1000 of a
proton size). The momentum uncertainty of a preon (of whatever 
mass) confined to a box of this size is about 200 GeV, which
is 50,000 times larger than the mass of the up-quark. 
Thus, the problem consists in reconciling the relatively 
small quark masses with the many orders of magnitude 
greater mass-energies arising from the preons' enormous momenta.

One way in which the mass from internal momentum can be cancelled
is to postulate an extremely strong force, which must be at 
least $10^5$ times stronger than the strong
interaction. It is somewhat unwelcome because it would add a 
considerable complication to the Standard Model, which already has too
many arbitrary parameters. However, with such a hyperforce, 
the preons would be so tightly bound inside a quark that the 
energy contribution from their large momentum 
would be cancelled by their large binding energy. This approach 
is quite promising, and that is why we adhere to 
it in this paper. 

We have also assumed that infinite energies
are not accessible in nature. Then, since it is an experimental
fact that energy usually increases as distance decreases, 
we hypothesize  that energy of any field on small 
scales, after reaching a maximum,
decays to zero at the origin. With this approach, one can 
use classical (intrinsically anomaly-free) potentials 
on small scales without being troubled by infinite energies or
anomalies.


\end{document}